\documentclass[aps,prc,twocolumn,showpacs,superscriptaddress]{revtex4-1}

\usepackage{subfigure}
\usepackage{graphicx}
\usepackage{float}
\usepackage{array}
\usepackage{graphicx}
\usepackage{amsmath,amsthm,amssymb}
\usepackage{color}
\usepackage[bookmarks,
                bookmarksopen = true,
                bookmarksnumbered = true,
                linktocpage,
                colorlinks = true,
                linkcolor = blue,
                urlcolor  = blue,
                citecolor = blue,
                anchorcolor = green,
                hyperindex = true,
                hyperfigures]
                {hyperref}

\def\lsim{\,\raise0.3ex\hbox{$<$\kern-0.75em\raise-1.1ex\hbox{$\sim$}}\,}
\def\gsim{\,\raise0.3ex\hbox{$>$\kern-0.75em\raise-1.1ex\hbox{$\sim$}}\,}

\newcommand{\nn}{\nonumber\\}
\newcommand{\ba}{\begin{eqnarray}}
\newcommand{\ea}{\end{eqnarray}}
\newcommand{\la}[1]{\label{#1}}

\begin{document}

\title{Event-by-event jet anisotropy and hard-soft tomography of the quark-gluon plasma}

\author{Yayun He}
\affiliation{Key Laboratory of Quark and Lepton Physics (MOE) and Institute of Particle Physics, Central China Normal University, Wuhan 430079, China}
\affiliation{Guangdong Provincial Key Laboratory of Nuclear Science, Institute of Quantum Matter, South China Normal University, Guangzhou 510006, China}
\affiliation{Guangdong-Hong Kong Joint Laboratory of Quantum Matter, Southern Nuclear Science Computing Center, South China Normal University, Guangzhou 510006, China}
\author{Wei Chen}
\affiliation{School of Nuclear Science and Technology, University of Chinese Academy of Sciences, Beijing 100049, China}
\author{Tan Luo}
\affiliation{Instituto Galego de F\'isica de Altas Enerx\'ias IGFAE, Universidade de Santiago de Compostela, E-15782 Galicia-Spain}
\author{Shanshan Cao}
\email{shanshan.cao@sdu.edu.cn}
\affiliation{Institute of Frontier and Interdisciplinary Science, Shandong University, Qingdao, Shandong 266237, China}
\author{Long-Gang Pang}
\email{lgpang@mail.ccnu.edu.cn}
\affiliation{Key Laboratory of Quark and Lepton Physics (MOE) and Institute of Particle Physics, Central China Normal University, Wuhan 430079, China}
\author{Xin-Nian Wang}
\email[]{xnwang@lbl.gov}
\affiliation{Key Laboratory of Quark and Lepton Physics (MOE) and Institute of Particle Physics, Central China Normal University, Wuhan 430079, China}
\affiliation{Nuclear Science Division Mailstop 70R0319,  Lawrence Berkeley National Laboratory, Berkeley, California 94720}
\email{current address}

\begin{abstract}
     Suppression of jet spectra or jet quenching in high-energy heavy-ion collisions is caused by jet energy loss in the dense medium. The azimuthal anisotropy of jet energy loss in non-central heavy-ion collisions can lead to jet anisotropy which in turn can provide insight into the path-length dependence of jet quenching. This is investigated within the Linear Boltzmann Transport (LBT) model which simulates both elastic scattering and medium-induced gluon radiation based on perturbative QCD for jet shower and medium recoil partons as well as radiated gluons as they propagate through the quark-gluon plasma (QGP). The dynamical evolution of the QGP in each event of heavy-ion collisions is provided by the (3+1)D CLVisc hydrodynamic model with fully fluctuating initial conditions. This framework has been shown to describe the suppression of single inclusive jet spectra well. We calculate in this study the elliptic ($v_{2}^{\rm jet}$) and triangular ($v_{3}^{\rm jet}$) anisotropy coefficients of the single inclusive jet spectra in Pb+Pb collisions at the LHC energies. We investigate the colliding energy, centrality, jet transverse momentum dependence of the jet anisotropy, as well as their event-by-event correlation with the flow coefficients of the soft bulk hadrons. An approximate linear correlation between jet and bulk $v_2$ is found. Effect of the bulk $v_n$ fluctuation on $v_n^{\rm jet}$ is found negligible. The jet-induced medium excitation, which is influenced by radial flow, is shown to enhance $v_{2}^{\rm jet}$ and the enhancement increases with the jet cone size. The jet elliptic anisotropy $v_{2}^{\rm jet}$ is also found to be slightly enhanced by the shear viscosity of the bulk medium in comparison to the LBT results when jets propagate through an ideal hydrodynamic QGP medium.
\end{abstract}


\maketitle

\section{Introduction}
\label{intro}

Jet quenching or the suppression of energetic particles from hard processes has been very successful in probing properties of the hot and dense matter called quark-gluon plasma (QGP) created in heavy-ion collisions. This phenomenon is caused by energy loss and transverse momentum broadening of a propagating parton due to multiple scattering and induced gluon bremsstrahlung when it traverses the dense medium \cite{Gyulassy:1990ye,Wang:1991xy,Gyulassy:2003mc,Majumder:2010qh,Mehtar-Tani:2013pia, Qin:2015srf,Cao:2020wlm}. Within perturbative QCD (pQCD), one can calculate the parton energy loss and transverse momentum broadening \cite{Baier:1996kr,Baier:1996sk,Zakharov:1996fv,Gyulassy:2000er,Wiedemann:2000tf,Wang:2001ifa,Arnold:2002ja}. They are found to be directly proportional to the jet transport coefficient $\hat q$ which is defined as the transverse momentum broadening squared per unit length of propagation and can be related to the local gluon density distribution of the medium \cite{CasalderreySolana:2007sw,Liang:2008vz,Zhang:2019toi}. Experimental measurements of jet quenching and phenomenological extraction of the jet transport coefficient can provide some of the important properties of QGP \cite{Chen:2010te,Burke:2013yra,Xie:2019oxg,Feal:2019xfl,JETSCAPE:2021ehl}.

Given the jet transport coefficient, the total transverse momentum broadening squared of a parton will be proportional to the total length of the propagation. The total radiative energy loss is, however, proportional to the length squared due to the non-Abelian Landau-Pomeranchuck-Migdal interference in the gluon bremsstrahlung induced by multiple scattering inside the medium \cite{Baier:1996kr,Baier:1996sk}. This unique length dependence of the radiative parton energy loss in pQCD will give rise to a unique system size dependence of the jet quenching phenomenon in high-energy heavy-ion collisions. In non-central heavy-ion collisions, the averaged path-length and total parton energy loss will depend on the azimuthal angle of the jet propagation relative to the reaction plane. Such an azimuthal angle dependence of the total energy loss was predicted \cite{Wang:2000fq} to give rise to the azimtuhal angle dependence or azimtuhal anisotropy of high transverse momentum jet and hadron spectra in non-central heavy-ion collisions which is very similar to the anisotropy of soft hadrons generated by the collective expansion and flow of the dense medium \cite{Ollitrault:1992bk}. The study of jet azimuthal anisotropy therefore will provide us additional information about jet propagation, the geometrical and dynamic properties of the dense medium in heavy-ion collisions.

Because of the rapid decrease of gluon number density or the local jet transport coefficient with time due to longitudinal and transverse expansion, the azimuthal anisotropy of the averaged total parton energy loss and the final hadron spectra are reduced \cite{Gyulassy:2000gk,Gyulassy:2001kr} from that of a simple 1-dimensional Bjorken system \cite{Wang:2000fq,Wang:2003mm}. The observed $v_2$ of charged hadrons at large transverse momentum $p_\mathrm{T}$ is larger than simple jet quenching model calculations that take into account of the geometry and hydrodynamic expansion of the dense medium in non-central heavy-ion collisions, especially at intermediate $p_\mathrm{T} \sim 4$-$10$~GeV/$c$. The hadron anisotropy below $p_\mathrm{T}<2$~GeV/$c$ is shown to come mostly from collective flow of the dense medium and can be described well by viscous hydrodynamic models \cite{Shen:2020gef,Heinz:2013th,Gale:2013da,deSouza:2015ena}. Simultaneous description of single inclusive hadron suppression and anisotropies have been shown to provide stronger constraint on jet transport dynamics, initial state of the QGP and the event-by-event fluctuation of the bulk medium \cite{Betz:2011tu,Zigic:2018ovr,Zigic:2019sth,Andres:2019eus,Noronha-Hostler:2016eow}. Some exotic mechanisms such as interaction through magnetic monopoles in QCD \cite{Xu:2014tda,Xu:2015bbz,Shi:2019nyp} 
and a singular behavior of the temperature dependence of the jet transport coefficient near the QCD phase transition 
have also been proposed to resolve the $v_2$ puzzle. However, it is quite likely that a mundane physics mechanism could be the culprit. It is known that recombination of jet shower and medium partons can lead to enhancement of protons and kaons and their anisotropic flow in intermediate $p_\mathrm{T}$ in heavy-ion collisions \cite{Fries:2003vb,Fries:2003kq,Greco:2003xt,Greco:2003mm,Greco:2003vf,Molnar:2003ff,Hwa:2008um} as well as in p+A collisions \cite{Hwa:2004zd,Zhao:2019ehg}. Such a mechanism can also provide a consistent description of the suppression and $v_2$ of the single inclusive hadron spectra in the intermediate $p_\mathrm{T}$. Indeed, parton energy loss and recombination in the hadronization can describe well  both the suppression factor and $v_2$ of charmed $D$ and beauty $B$ mesons \cite{Cao:2021ces,Dong:2019byy,Xing:2019xae} as well as the light hadrons in heavy-ion collisions \cite{Zhao:2021vmu}.

Since jets are reconstructed from clusters of hadrons within a given jet-cone, their energies are directly related to the parton energies before hadronization. They are less likely influenced by the non-perturbative hadronization processes which only contribute to the jet energy about 1~GeV within a cone-size of $R=1$~\cite{Ellis:1990ek}. The jet suppression and jet anisotropy can therefore be directly related to parton transport, geometry and dynamical evolution of the QGP in heavy-ion collisions. This will be the focus of our study in this paper. We will employ the Linear Boltzmann Transport (LBT) model \cite{Wang:2013cia,He:2015pra,Cao:2016gvr} together with the (3+1)-dimensional CCNU (Central China Normal University) - LBNL (Lawrence Berkeley National Laboratory)  viscous (CLVisc) hydrodynamic model \cite{Pang:2014ipa,Pang:2012he,Pang:2018zzo} for bulk medium evolution to investigate event-by-event jet anisotropy in Pb+Pb collisions at the Large Hadron Collider (LHC) energies. In particular, we will study the correlation between anisotropies of hard jets and soft bulk hadrons, effects of the event-by-event fluctuation of the bulk hadron spectra, jet-induced medium response and viscosity of the bulk medium. The same framework has been shown to describe the suppression of single inclusive jet spectra \cite{He:2018xjv} as well as $\gamma/Z$-jet correlation \cite{Luo:2018pto,Zhang:2018urd} in Pb+Pb collisions at the LHC energies.

The remainder of this paper will be organized as follows. We will start with a brief review of the LBT model for jet transport in Sec.~\ref{sec:lbt} and the CLVisc hydrodynamic model for the bulk evolution in Sec.~\ref{sec:hydro}. After that, we will present and discuss our results on the elliptic flow coefficient ($v_2$) of single inclusive jets in Sec.~\ref{sec:v2} and triangular flow coefficient ($v_3$) in Sec.~\ref{sec:v3}. In Sec.~\ref{sec:mediumResponse}, effects of jet-induced medium response on jet anisotropic coefficients and their dependence on the jet-cone size will be explored in detail. The effect of the viscosity of the bulk medium on the final jet elliptic anisotropy $v_2$ will be discussed in Sec.~\ref{sec:viscosity}.  A summary will be given in Sec.~\ref{sec:summary}.

\section{The Linear Boltzmann Transport model}
\label{sec:lbt}
The LBT model is based on the Boltzmann transport equation that includes both elastic and inelastic scatterings of jet shower partons as well as recoil medium partons inside the QGP medium~\cite{Wang:2013cia,He:2015pra,Cao:2016gvr}: 
\begin{eqnarray}
p_a\cdot\partial f_a&=&\int \sum_{b c d } \prod_{i=b,c,d}\frac{d^3p_i}{2E_i(2\pi)^3} (f_cf_d-f_af_b)|{\cal M}_{ab\rightarrow cd}|^2
\nn && \hspace{-0.5in}\times \frac{\gamma_b}{2}
S_2(\hat s,\hat t,\hat u)(2\pi)^4\delta^4(p_a\!+\!p_b\!-\!p_c\!-\!p_d)+ {\rm inelastic},
\label{bteq}
\end{eqnarray}
where $p_{i}\, (i = a, b, c, d)$ is the four momentum of parton $i$ whose phase-space distribution function is $f_{i} = (2 \pi)^3 \delta^{3} (\vec{p} - \vec{p}_{i}) \delta^{3} (\vec{x} - \vec{x}_{i} - \vec{v}_{i} t) $ for jet shower and recoil partons $(i = a, c)$, and $f_{i} = 1/(e^{p_{i}\cdot u / T} \pm 1)$ for thermal medium partons $(i = b, d)$ that are assumed in local equilibrium with temperature ${T}$ and flow velocity $u$ and ``$+$ ($-$)" for quarks (gluons). Note that this equation includes both a gain term ($f_c f_d$) and a loss term ($-f_a f_b$). Here, the quantum effects of Pauli exclusion and Bose enhancement in the transport equation are assumed small and neglected. The summation over $b$, $c$ and $d$ takes into account all possible elastic processes $a + b \rightarrow c + d$ in which the scattering amplitude $|{\cal M}_{ab \rightarrow cd}|$ is calculated in the leading-order pQCD. The factor $\gamma_{b}$ represents the color and spin degeneracy of parton $b$. To regularize the collinear divergence in the scattering amplitude, a Lorentz-invariant double-step function is adopted as follows,
\begin{equation}
S_2(\hat s, \hat t, \hat u) = \theta(\hat s\ge 2\mu_{D}^2)\theta(-\hat s+\mu_{D}^2\le \hat t\le -\mu_{D}^2),
\end{equation}
where $\hat s$, $\hat t$, and $\hat u$ are the Mandelstam variables, and $\mu_{D}^2 = \frac{3}{2}g^2 T^2$ is the Debye screening mass taking into account both quark (three light flavors) and gluon degrees of freedom. A fixed strong coupling constant $\alpha_\mathrm{s} = g^2/4\pi$ is used in the present work, which is tuned to describe experimental data on single inclusive jet spectra.

The inelastic part in Eq.~\eqref{bteq} includes gluon radiation induced by the elastic scattering processes discussed above. The induced gluon emission rate $\Gamma^{\rm inel}_{a}$ is taken from the higher-twist calculation~\cite{Guo:2000nz,Wang:2001ifa,Majumder:2009ge,Zhang:2003wk},
\ba \la{induced}
\frac{d\Gamma_{a}^{\rm inel}}{dzdk_\perp^2}=\frac{6\alpha_\mathrm{s} P_a(z) k_\perp^4}{\pi (k_\perp^2+z^2m^2)^4} \frac{p\cdot u}{p_0}\hat{q}_a(x)\sin^2 \frac{\tau-\tau_i}{2\tau_f},
\label{eq:HTspectra}
\ea
where $P_{a}(z)$ is the splitting function for parton $a$ (with mass $m$) to radiate a gluon with energy fraction $z$ and transverse momentum $k_{\perp}$. The infrared divergence of the splitting function is also regularized by the Debye screening mass $\mu_{D}$ as the lower cut-off energy of the radiated gluon. In the sine term, $\tau_{f}=2 p_0 z (1-z) /(k_\perp^2 + z^2 m^2)$ is the formation time of the emitted gluon, and $\tau_{i}$ is the production time of the present parton, i.e., the time of the previous gluon emission. The jet transport coefficient $\hat{q}_a(x)$ represents the transverse momentum transfer squared per unit length/time due to elastic scatterings and is evaluated in the local comoving frame of the QGP medium as
\begin{equation}
\hat{q}_{a}(x)=\sum_{bcd}\rho_{b}(x)\int d\hat t q_\perp^2 \frac{d\sigma_{ab\rightarrow cd}}{d\hat t},
\label{eq-qhat}
\end{equation}
where $\rho_b(x)$ is the parton density with the color and spin degeneracy included.

Given the elastic and inelastic scattering rates
\begin{equation}
\Gamma_a^{\rm el} = \frac{p\cdot u}{p_0}\sum_{bcd} \rho_b(x)\sigma_{ab\rightarrow cd},
\label{eq:rateEl}
\end{equation}
\begin{equation}
\Gamma_a^{\rm inel}=\frac{1}{1+\delta_g^a}\int dz dk_\perp^2 \frac{d\Gamma_a^{\rm inel}}{dzdk_\perp^2},
\label{eq:rateInel}
\end{equation}
from Eqs.~(\ref{bteq}) and (\ref{eq:HTspectra}), 
the elastic and inelastic scattering probabilities of parton $a$ within a time step $\Delta \tau$ in the LBT simulations are calculated as 
\begin{equation}
P^a_{\rm el}=1-\text{exp}[- \Delta\tau \Gamma_a^{\rm el}(x)]
\label{eq:probEl}
\end{equation}
and 
\begin{equation}
P^a_\mathrm{inel}=1-\exp[-\Delta\tau \Gamma_a^{\rm inel}(x)]
\label{eq:probInel}
\end{equation}
respectively, where we assume the numbers of both elastic and inelastic scatterings during $\Delta \tau$ obey Poisson distributions, whose average values are $\Delta\tau \Gamma_a^{\rm el}$ and $\Delta\tau \Gamma_a^{\rm inel}$, respectively. The latter may also be interpreted as the average number of emitted gluons $\langle N^a_g \rangle$ during this time interval. Therefore, Eq.~(\ref{eq:probEl}) / (\ref{eq:probInel}) gives the probability for at least one elastic/inelastic scattering during this time interval  $\Delta \tau$.

The total scattering probability is given by 
\begin{align}
P^a_\mathrm{tot}=P^a_\mathrm{el}(1-P^a_\mathrm{inel}) +P^a_\mathrm{inel},
\end{align}
where the first term on the right-hand side is for pure elastic processes without induced gluon emission, and the second term is for inelastic processes with at least one gluon emission. Based on the total probability, we first decide whether a given parton scatters with the QGP during this $\Delta \tau$. If it scatters, the ratio between these two terms is used to determine whether the scattering is pure elastic or inelastic. In either case, an elastic scattering will be sampled first, whose specific channel is determined using the branching ratio $\Gamma_{a+b \rightarrow c+d}^{\rm el} / \Gamma_a^{\rm el}$. With a selected channel, the energies and momenta of partons $b$, $c$ and $d$ are then sampled with the differential rate given by the first part of Eq.~(\ref{bteq}). In case of inelastic scattering, the number of emitted gluons $n$ will be first decided according to a Poission distribution with the mean value $\langle N^a_g \rangle$. The energy-momentum of each gluon is then sampled with the spectra given by Eq.~(\ref{eq:HTspectra}). In the end, the energies and momenta of these $n$ gluons will be adjusted together with those of $c$ and $d$ such that the $2\rightarrow 2+n$ process respects the energy-momentum conservation. 

In the LBT model, we track all partons involved in the scatterings. We define $c$ or $d$ with a larger energy as the jet shower parton, while the other as the recoil parton. Medium induced gluons are also tracked as jet shower partons. Parton $b$ which is scattered out of the thermal medium background is denoted as a ``negative" parton from the back-reaction and is allowed to go through further scattering in LBT. It is essentially a particle hole left in the medium when the original thermal parton is scattered out. Thus, its four-momentum will be subtracted from the reconstructed jets in our analysis of simulated events from LBT. Both recoil and ``negative'' (back-reaction) partons are considered as jet-induced medium excitation, or medium response to jet propagation, whose importance has been verified in many jet observables within the LBT model~\cite{Luo:2018pto,He:2018xjv}.

The term ``linear" in LBT denotes a linear approximation adopted in this model, where we only consider jet/recoil/``negative" parton interaction with the medium, but not among themselves, assuming their number density negligible compared to the thermal partons inside the QGP. A full calculation that includes such non-linear interactions can be realized within a coupled LBT-hydrodynamic (CoLBT-hydro) model that has been developed in Refs.~\cite{Chen:2017zte,Chen:2020tbl,Yang:2021iib}. Such a coupled approach is important to describing properties of the jet-induced medium response in detail. But the effect of interaction among recoil and soft radiated gluons beyond the linear approximation is negligible on the energy of reconstructed jets and the final jet spectra.

\section{CLVisc Hydrodynamics for bulk medium evolution}
\label{sec:hydro}

To take into account the evolution of the QGP in heavy-ion collisions in the LBT model, we use the space-time profile of the bulk medium from the CLVisc (3+1)D viscous hydrodynamic model~\cite{Pang:2014ipa,Pang:2012he,Pang:2018zzo}. CLVisc parallelizes Kurganov-Tadmor algorithm \cite{KURGANOV2000241} to solve the hydrodynamic equation for the bulk medium and Cooper-Frye particlization on GPU, using Open Computing Language (OpenCL). Parallelized with massive amount of processing elements on GPUs and Single Instruction Multiple Data (SIMD) vector operations on modern CPUs, CLVisc brings about the best performance increase so far to (3+1)D hydrodynamics on heterogeneous computing devices and provides the event-by-event space-time hydrodynamic profiles for simulations of jet transport within LBT model in this study.

The dynamical evolution of the locally thermalized system in heavy-ion collisions is described by relativistic hydrodynamic equations,
\begin{equation}
    \nabla_{\mu} T^{\mu\nu} = 0,
\end{equation}
where $\nabla_{\mu}$ is the covariant derivative operator, 
$T^{\mu\nu} = (\epsilon+P)u^{\mu}u^{\nu} - P g^{\mu\nu} + \pi^{\mu\nu}$ is the energy-momentum stress tensor, in which $\epsilon$ and $P$ are the energy density and pressure in the co-moving frame of the fluid,
$u^{\mu}$ is the relativistic fluid four-velocity,
$g^{\mu\nu} = \rm{diag}(1, -1, -1, -\tau^2)$ is the metric tensor in the Milne $(\tau, x, y, \eta_s)$ coordinates and $\pi^{\mu\nu}$ is the shear-stress tensor which will depend on the bulk transport coefficients. In the case of an ideal hydrodynamics, this term is set to zero. To solve this group of time-dependent partial differential equations, one needs the
energy-momentum tensor $T^{\mu\nu}$ at the initial time $\tau_0$ and the equation of state (EoS) $P=P(\varepsilon)$.

The initial condition for the energy-momentum density distributions for event-by-event CLVisc hydrodynamic simulations in this study are obtained from A Multi-Phase Transport (AMPT) model~\cite{Lin:2004en,Pang:2012he} with a Gaussian smearing,
\begin{equation}
  \begin{aligned}
  T^{\mu\nu} &(\tau_{0},x,y,\eta_{s}) = K\sum_{i}
  \frac{p^{\mu}_{i}p^{\nu}_{i}}{p^{\tau}_{i}}\frac{1}{\tau_{0}\sqrt{2\pi\sigma_{\eta_{s}}^{2}}}\frac{1}{2\pi\sigma_{r}^{2}}\\
     	  &\hspace{-0.1in} \times \exp \left[-\frac{(x-x_{i})^{2}+(y-y_{i})^{2}}{2\sigma_{r}^{2}} - \frac{(\eta_{s}-\eta_{i s})^{2}}{2\sigma_{\eta_{s}}^{2}}\right],
  \end{aligned}
  \label{eq:Pmu}
\end{equation}
where $p^{\tau}_{i}=m_{i\rm T}\cosh(Y_{i}-\eta_{i s})$ and $p^{\eta}_{i}=m_{i\rm T}\sinh(Y_{i}-\eta_{i s})/\tau_{0}$ with $m_{i\rm T}=\sqrt{p_{ix}^2+p_{iy}^2+m^2}$. The summation runs over all partons $(i)$ produced in the AMPT model simulations.  We have chosen $\sigma_{r}=0.6$~fm and $\sigma_{\eta_{s}}=0.6$ in our calculations which provide a reasonable description of soft hadron observables \cite{Pang:2018zzo}. The transverse mass $m_{\rm T}$, rapidity $Y$ and spatial rapidity $\eta_{s}$ are calculated from parton's $4$-momenta and spatial coordinates.  Note that there is no Bjorken scaling in the above initial condition because of early parton cascade in AMPT model. The scale factor $K$ and the initial time $\tau_{0}$ are two parameters that can be adjusted to fit the experimental data on the central rapidity density of produced hadrons.

For most calculations in this study, we use the ideal version of CLVisc with a parametrized equation of state (EoS) s95p-v1~\cite{Huovinen:2009yb} to obtain the hydrodynamic evolution of the bulk medium. We will discuss the effect of shear viscosity on jet quenching and anisotropy at the end of this paper. In each centrality bin, we simulate 200 events of hydrodynamic evolution of the dense medium in heavy-ion collisions in order to include the effect of event-by-event fluctuations on jet transport. To improve the statistics of high-$p_{\rm T}$ jets, we divide the initial transverse momentum transfer $p_{{\rm T}i}$ in PYTHIA 8 simulations into multiple bins
and generate 10000 sets of initial jet showers from PYTHIA 8 in each $p_{{\rm T}i}$  bin for each of the above-mentioned hydrodynamic profile. The total number of jet events for each $p_{{\rm T} i}$ bin is therefore $N_{\rm event}=200\times 10000$ for each centrality bin under investigation. This is also the total number of events we use for simulating $p+p$ collisions in each $p_{{\rm T}i}$ bin for the calculation of the single inclusive jet suppression factor \cite{He:2018xjv}.  The jet cross sections in different $p_{{\rm T}i}$ bins of initial hard scatterings will be used as weights in calculating the final jet spectra in both $p+p$ and A+A collisions.

\begin{figure}[tbp]
    \centerline{\includegraphics[width=11cm]{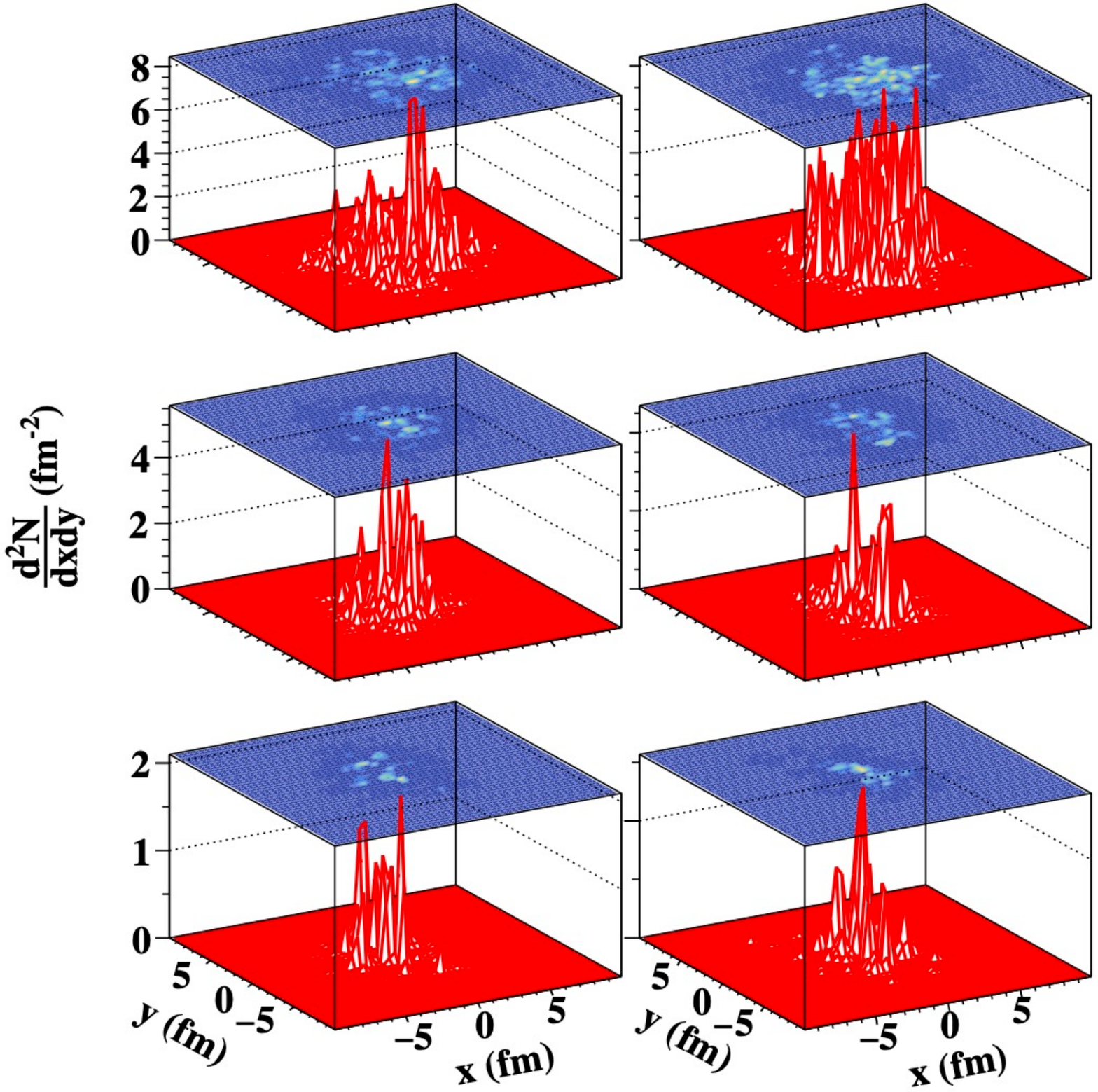}}
    \centerline{\includegraphics[width=11cm]{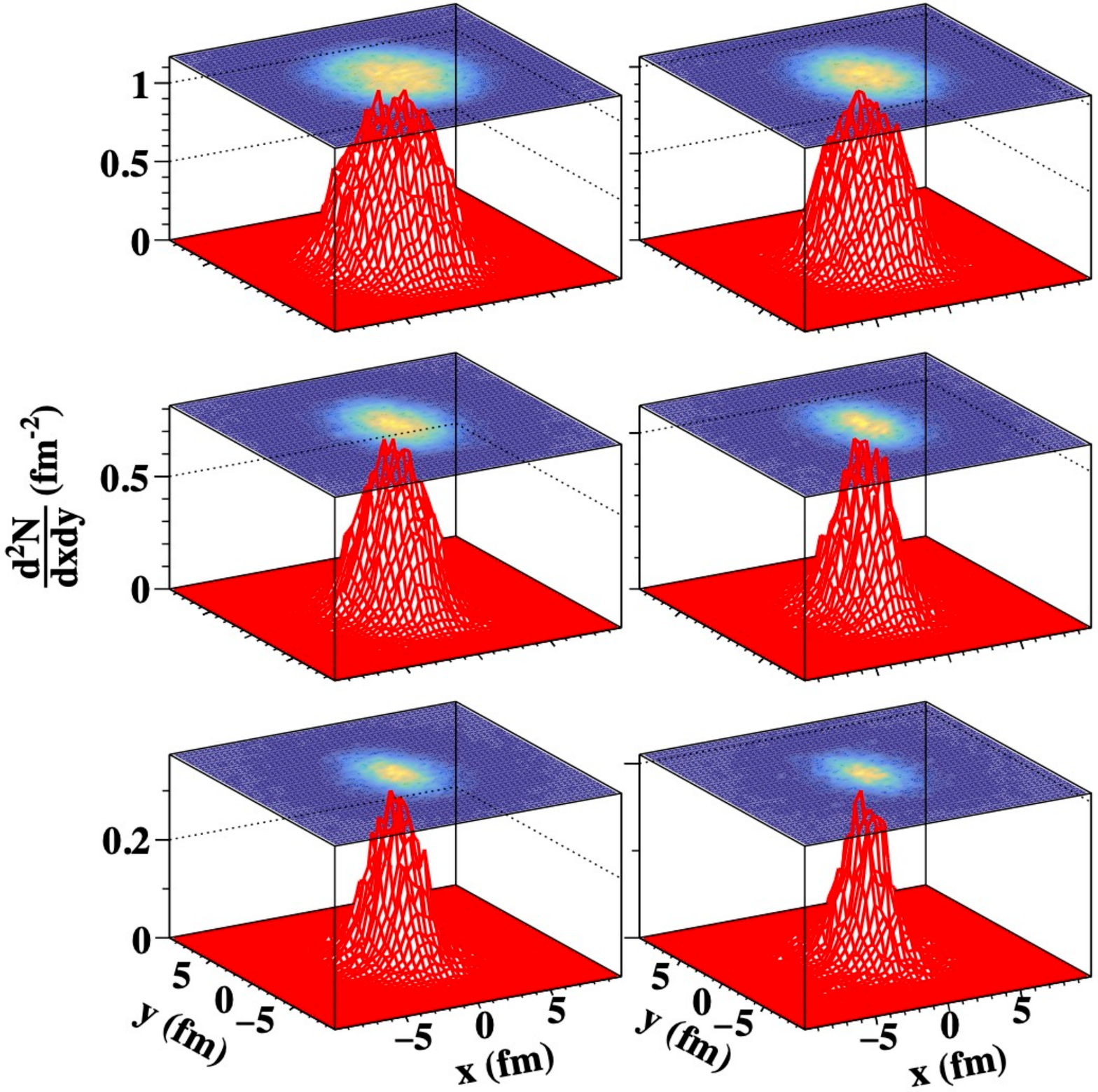}}
    \caption{(Color online) Distributions of initial jet production positions in the transverse plane of Pb+Pb collisions at $\sqrt{s} = 5.02$~TeV in the centrality bins $5 - 10 \%, 10 - 20 \%, 20 - 30 \%, 30 - 40 \%, 40 - 50 \%$ and $50 - 60 \%$ (from left to right and top to bottom) for one hydrodynamic event (upper panel) and averaged over 200 hydrodynamic events (lower panel).}
    \label{geo_5020}
\end{figure}

For the spatial distribution of the initial jet production vertices, we also use the AMPT model that employs the HIJING model~\cite{Wang:1991hta,Gyulassy:1994ew} to generate minijets according to the Glauber model of nuclear collisions with the Woods-Saxon nuclear distribution. The geometrical distribution of the initial jets in the transverse plane in each $p_{{\rm T}i}$ bin is sampled according to this initial minijet distribution in each AMPT event, as shown in Fig.~\ref{geo_5020}. The same AMPT event also provides the initial condition for the energy-momentum density distribution for the CLVisc hydrodynamic simulation of the space-time evolution of the bulk medium. The centrality classes of heavy-ion collisions are defined according to the initial parton multiplicity distribution, and the averaged number of participant nucleons $\langle N_{\rm part}\rangle$ in each centrality class is computed accordingly. 


\begin{figure}[tbp]
    \includegraphics[width=8cm]{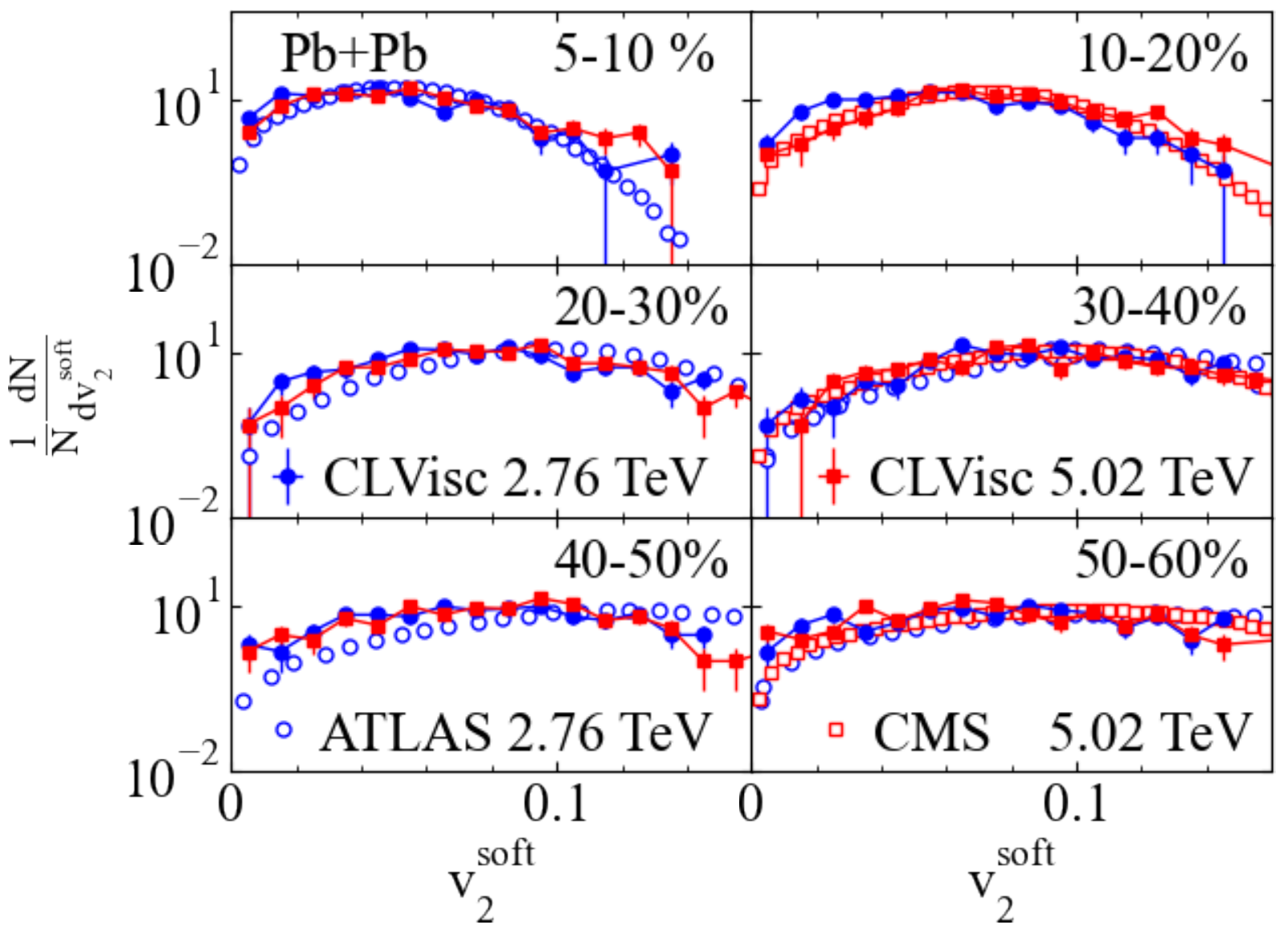}
    \caption{(Color online) Distributions of soft hadron $v^\mathrm{soft}_2$ from 200 hydro events calculated with CLVisc model in Pb+Pb collisions at $\sqrt{s} = 2.76$~TeV (closed blue circle) and $5.02$~TeV (closed red square) in each centrality bin $5 - 10 \%, 10 - 20 \%, 20 - 30 \%, 30 - 40 \%, 40 - 50 \%$ and $50 - 60 \%$ as compared to data from the ATLAS experiments in Pb+Pb collisions at $\sqrt{s} = 2.76$ TeV (open blue circle) in centrality bins $5 - 10 \%, 20 - 25 \%, 30 - 35 \%, 40 - 45 \%, 55 - 60 \%$~\cite{Aad:2013xma} and the CMS experiment at $\sqrt{s} = 5.02$ TeV (open red square) in centrality bins $15 - 20 \%, 30 - 35 \%, 55 - 60 \%$~\cite{Sirunyan:2017fts}.} 
    \label{v2soft}
\end{figure}

\begin{figure}[tbp]
    \includegraphics[width=8cm]{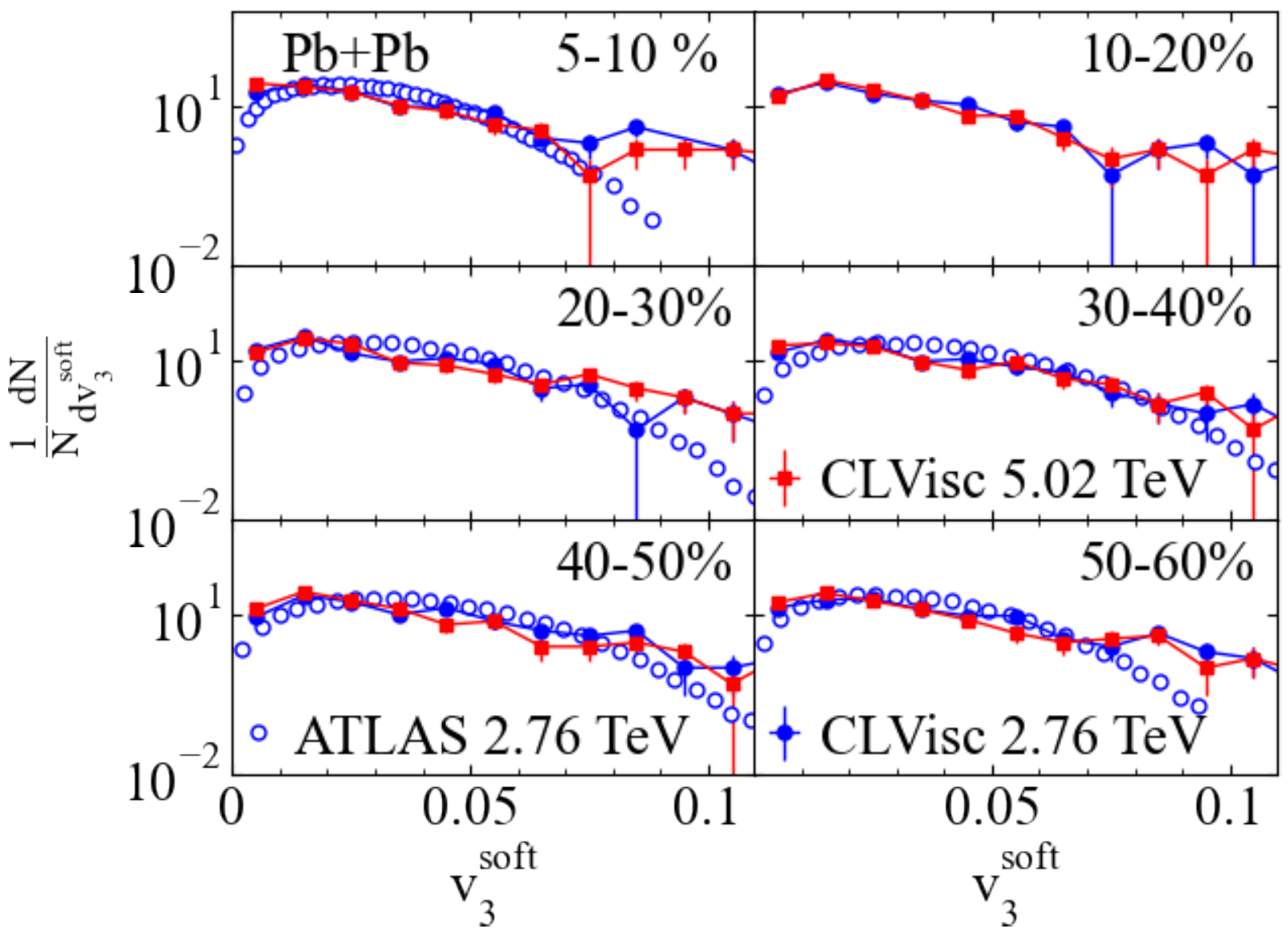}
    \caption{(Color online) Distributions of soft hadron $v^\mathrm{soft}_3$ from 200 hydro events calculated with CLVisc model in Pb+Pb collisions at $\sqrt{s} = 2.76$~TeV (closed blue circle) and $5.02$~TeV (closed red square) in centrality bin $5 - 10 \%, 10 - 20 \%, 20 - 30 \%, 30 - 40 \%, 40 - 50 \%$ and $50 - 60 \%$ as compared to the ATLAS experimental data at Pb+Pb $\sqrt{s} = 2.76$ TeV (open blue circle) in centrality bins $5 - 10 \%, 20 - 25 \%, 30 - 35 \%, 40 - 45 \%, 55 - 60 \%$~\cite{Aad:2013xma}.} 
    \label{v3soft}
\end{figure}

\begin{table} [tbp]
\centering
$\langle v^\mathrm{soft}_2\rangle \pm \delta v_2^{\rm soft}$ \\
\begin{tabular}{ || c| c| c  || }
\hline
 & 2.76 TeV & 5.02 TeV  \\ [0.5 ex]
\hline \hline
$5-10 \%$ & $0.047\pm 0.007$ & $0.054\pm 0.008$ \\
$10-20\%$ & $0.060\pm 0.008$ & $0.076\pm 0.007$ \\
$20-30\%$ & $0.076\pm 0.008$ & $0.086\pm 0.008$ \\
$30-40\%$ & $0.089\pm 0.008$ & $0.095\pm 0.009$ \\
$40-50\%$ & $0.079\pm 0.008$ & $0.086\pm 0.009$ \\
$50-60\%$ & $0.078\pm 0.009$ & $0.078\pm 0.009$ \\
[1ex]
\hline
\end{tabular}
\caption{The mean values and standard deviations of soft hadron $v^\mathrm{soft}_2$ in Pb+Pb collisions at $\sqrt{s} = 2.76$~TeV and 5.02~TeV in centrality bins $5 - 10 \%, 10 - 20 \%, 20 - 30 \%, 30 - 40 \%, 40 - 50 \%$ and $50 - 60 \%$ from the CLVisc model.}
\label {table:v2_mean_std}
\end{table}

\begin{table} [tbp]
\centering
$\langle v^\mathrm{soft}_3\rangle \pm \delta v_3^{\rm soft}$ \\
\begin{tabular}{ || c| c| c  || }
\hline
 & 2.76 TeV & 5.02 TeV  \\ [0.5 ex]
\hline \hline
$5-10 \%$ & $0.031\pm 0.007$ & $0.027\pm 0.007$ \\
$10-20\%$ & $0.031\pm 0.007$ & $0.029\pm 0.007$ \\
$20-30\%$ & $0.032\pm 0.007$ & $0.035\pm 0.008$ \\
$30-40\%$ & $0.034\pm 0.007$ & $0.035\pm 0.008$ \\
$40-50\%$ & $0.038\pm 0.007$ & $0.034\pm 0.008$ \\
$50-60\%$ & $0.035\pm 0.007$ & $0.032\pm 0.008$ \\
[1ex]
\hline
\end{tabular}
\caption{The mean value and standard deviation of soft hadron $v^\mathrm{soft}_3$ in Pb+Pb collisions at $\sqrt{s} = 2.76$~TeV and 5.02~TeV in centrality bins $5 - 10 \%, 10 - 20 \%, 20 - 30 \%, 30 - 40 \%, 40 - 50 \%$ and $50 - 60 \%$ from the CLVisc model.}
\label {table:v3_mean_std}
\end{table}

To demonstrate the event-by-event fluctuation of the bulk medium, we present the distributions of the elliptic and triangular flow coefficients of the final bulk hadrons from our event-by-event hydrodynamic calculations in Figs.~\ref{v2soft} and~\ref{v3soft}. Results for Pb+Pb collisions at both $\sqrt{s} = 2.76$~TeV and $5.02$~TeV are shown, and compared to available data in different centrality bins from ATLAS \cite{Aad:2013xma} and CMS experiments \cite{Sirunyan:2017fts}. A reasonable agreement between our hydrodynamic simulations and the experimental data is achieved except at very large values of $v_2^{\rm soft}$ and $v_3^{\rm soft}$ where statistics of the hydrodynamic simulations becomes very limited. To quantify these $v_n^{\rm soft}$ distributions,  we summarize the average values of $v_2^{\rm soft}$ and $v_3^{\rm soft}$ in Tabs.~\ref{table:v2_mean_std} and~\ref{table:v3_mean_std}, respectively, together with their corresponding fluctuations $\delta v_2^{\rm soft}$ and $\delta v_3^{\rm soft}$ for different centrality bins and colliding energies. One may observe from Tab.~\ref{table:v2_mean_std} that $\langle v_2^{\rm soft}\rangle$ first increases and then decreases as centrality increases. Within a given centrality bin, $\langle v_2^{\rm soft}\rangle $ in Pb+Pb collisions at $\sqrt{s} = 5.02$~TeV is larger than that at 2.76~TeV, except in very peripheral collisions. The event-by-event fluctuation of the elliptic flow $\delta v_2^{\rm soft}/v_2^{\rm soft}$ is around 10\%$\sim$15\%. In contrast, as shown in Tab.~\ref{table:v3_mean_std}, the bulk $\langle v_3^{\rm soft}\rangle$ has a quite weak dependence on centrality and colliding system under discussion, because it is mainly driven by event-by-event fluctuation instead of the average geometry of the bulk medium. The value of $\delta v_3^{\rm soft}/v_3^{\rm soft}$ is around 20\%$\sim$25\%.


\section{Single inclusive jet anisotropy $v^\mathrm{jet}_{2}$}
\label{sec:v2}

\subsection{LBT simulations and analyses}

To calculate jet spectra in realistic heavy-ion collisions, we first use PYTHIA 8 to generate initial jet shower partons in $p+p$ collisions at the corresponding colliding energy. In order to obtain sufficient statistics for our final results, we divide the $(0, 350)$~GeV/$c$ range of the transverse momentum transfer for the initial hard scatterings into seven equal bins, each with the width of 50~GeV/$c$. In each of these triggering $p_\mathrm{T}$ bins, 2 million events are simulated in total, with 10000 events in which jet shower partons propagate through each of the 200 hydrodynamic profiles that we obtain via the CLVisc model as discussed in the previous section. Using 200 hydrodynamic profiles per centrality bin allows one to take into account of the effects of the event-by-event fluctuations in the bulk medium on the final jet observables. The initial position of each jet production is sampled according to the distribution of hard scattering locations in the AMPT model for each of the hydro event, which we consistently use to determine the initial condition of the bulk evolution. In this work, we only consider the distribution of the transverse locations of these jet production points, while neglect their spread in the longitudinal direction around the highly Lorentz contracted disc of two overlapping nuclei at high colliding energies.

The formation time of each jet shower parton is set as $p_\mathrm{T}^2/2E$ with $p_\mathrm{T}$ and $E$ being its initial transverse momentum and energy, respectively. The parton is assumed to stream freely before its formation time is reached. After this formation time, as well as the starting time of the hydrodynamic evolution ($\tau_0 = 0.6$~fm), whichever comes later, we simulate the interaction between jet shower partons and the hydrodynamic medium using the LBT model. After jet partons exit the QGP medium, we neglect their interactions with the hadron gas in this work, considering that the gluon density in the hadronic gas is much lower and the effective jet transport coefficient $\hat q$ in a confined medium is much smaller \cite{Chen:2010te} than that in the QGP medium

In simulations of both $p+p$ and A+A collisions, we pass all final state partons to the FASTJET package~\cite{Cacciari:2011ma} to reconstruct jets using the anti-$\it{k}_\mathrm{T}$ algorithm with a given jet-cone size $R$. The FASTJET package used in this study has been modified such that the energy-momentum of the ``negative" partons from the LBT model is subtracted from jets in the reconstruction. When reconstructing jets via FASTJET, we also subtract the underlying event (UE) background using the scheme applied in experimental studies~\cite{Aad:2012vca}. One may first define the seed jet as a jet with at least one particle whose transverse energy is higher than 3~GeV, and with a leading particle whose transverse energy is more than four times of the average value per particle within the jet. Then the transverse energy density of the UE background is calculated over the whole area under investigation excluding these seed jets. In the end, this UE transverse energy within the transverse area of each jet is subtracted from the jet energy in both $p+p$ and A+A collisions. In LBT simulations, we only consider evolution of jet shower partons, recoil partons and ``negative" (back-reaction) partons, i.e., partons that directly participate in jet-medium interaction. Medium constituents that do not participate in this interaction are excluded for jet reconstruction. Therefore, the UE background in our analysis is very small compared to that in experimental analyses which also include soft hadrons from the QGP that are not directly correlated with the jet. Within this framework, we have verified in Ref.~\cite{He:2018xjv} that the jet spectra we obtain for $p+p$ collisions are consistent with the experimental data. With a fixed effective strong coupling constant $\alpha_\mathrm{s}=0.15$, LBT model provides a good description of the jet suppression factor $R_\mathrm{AA}$ in A+A collisions \cite{He:2018xjv}. It provides a reliable baseline for our further investigation of the azimuthal anisotropy of jets in the present work.

\begin{figure}[t]
    \includegraphics[width=8cm]{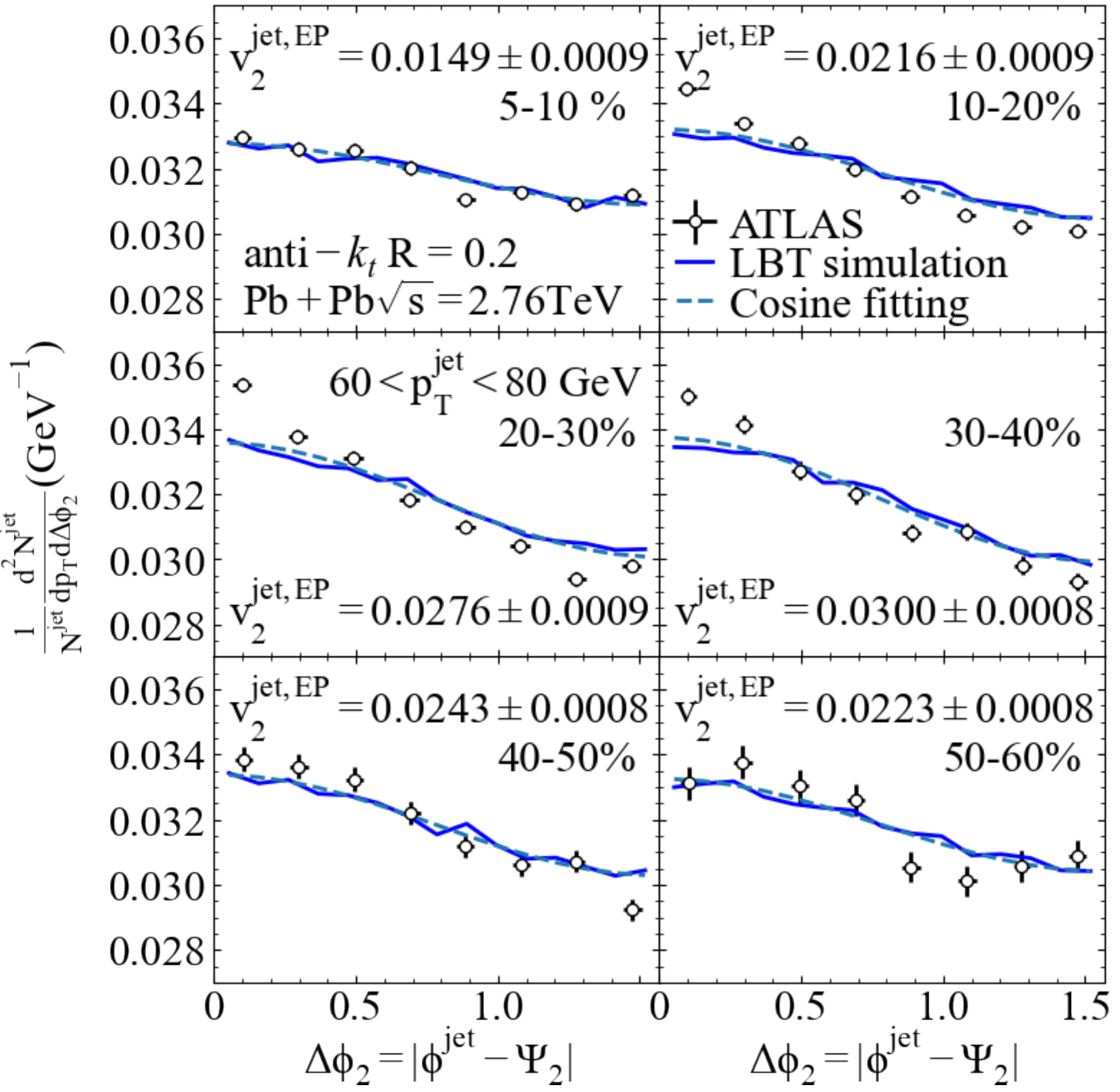}
    \caption{(Color online) Distributions of the difference of the jet azimuthal angle $\phi^\mathrm{jet}$ and the event plane angle $\Psi_{2}$ ($\Delta \phi_{2} = \phi^\mathrm{jet} - \Psi_{2}$) in different centrality bins of Pb+Pb collisions at $\sqrt{s} = 2.76$~TeV. Results calculated from the LBT model (blue solid lines) are compared to the ATLAS data~\cite{Aad:2013sla} (black open circle) and fitted with a cosine function (blue dash lines).}
    \label{jetphi2_2760}
\end{figure}

\subsection{Centrality dependence}

To extract the anisotropic flow coefficient of the single inclusive jet spectra in A+A collisions, we express their normalized azimuthal distribution as,
\begin{equation}
    \label{eq:v2EP}
    \frac{1}{N^\mathrm{jet}}\frac{dN^\mathrm{jet}}{d\Delta \phi_{n}} \propto 1 + 2 v_{n}^\mathrm{jet,EP} \cos(n\Delta \phi_{n}),
\end{equation}
 where $\Delta \phi_{n} = \phi^\mathrm{jet} - \Psi_{n}$ is the difference between the azimuthal angle of jets $\phi^\mathrm{jet}$ and the $n^\mathrm{th}$-order event plane angle $\Psi_{n}$, with $\Psi_{n}$ being defined via $\langle e^{in\phi^{\rm soft}}\rangle =v_n^{\rm soft}e^{in\Psi_{n}}$ for each hydrodynamic event. The superscript ``EP" in Eq.~(\ref{eq:v2EP}) denotes that this jet $v_n^{\rm jet}$ is defined via the event plane method. Shown in Fig.~\ref{jetphi2_2760} are the angular distributions for $n = 2$ in $\Delta \phi_{2}$ for different centrality bins of Pb+Pb collisions at $\sqrt{s} = 2.76$~TeV, as compared to the corresponding ATLAS data~\cite{Aad:2013sla}. Here, the jet radius is $R=0.2$ and the jet transverse momentum is in the interval $60 < p_\mathrm{T} < 80$~GeV/$c$. 
 
 We further fit these $\Delta \phi_{2}$ distributions with the function $2 / (\pi \Delta p_\mathrm{T}) ( 1 + 2 v_{2}^\mathrm{jet,EP} \cos(2 \Delta \phi_{2}))$, as shown by the solid blue lines in Fig.~\ref{jetphi2_2760}, from which we extract the elliptic flow coefficient $v^\mathrm{jet,EP}_2$. Here, the $2/\pi$ factor is introduced to normalize the jet spectrum within $\Delta\phi_2 \in [0,\pi/2]$. The corresponding values of $v^\mathrm{jet,EP}_2$ are indicated in the figure for different centralities. We observe that for central collisions (5-10\%), the jet elliptic coefficient is small due to the small geometric anisotropy of the average QGP profile. However, it is also non-zero due to the initial event-by-event geometrical fluctuation of the QGP. This $v^\mathrm{jet,EP}_2$ increases with the centrality as the geometry of the QGP fireballs becomes more anisotropic towards semi-peripheral collisions. However, for very large centralities, $v^\mathrm{jet,EP}_2$ decreases again, because the amount of jet energy loss is becoming smaller in these smaller systems, leading to a smaller $v^\mathrm{jet,EP}_2$.

\begin{figure}[tbp]
    \includegraphics[width=8cm]{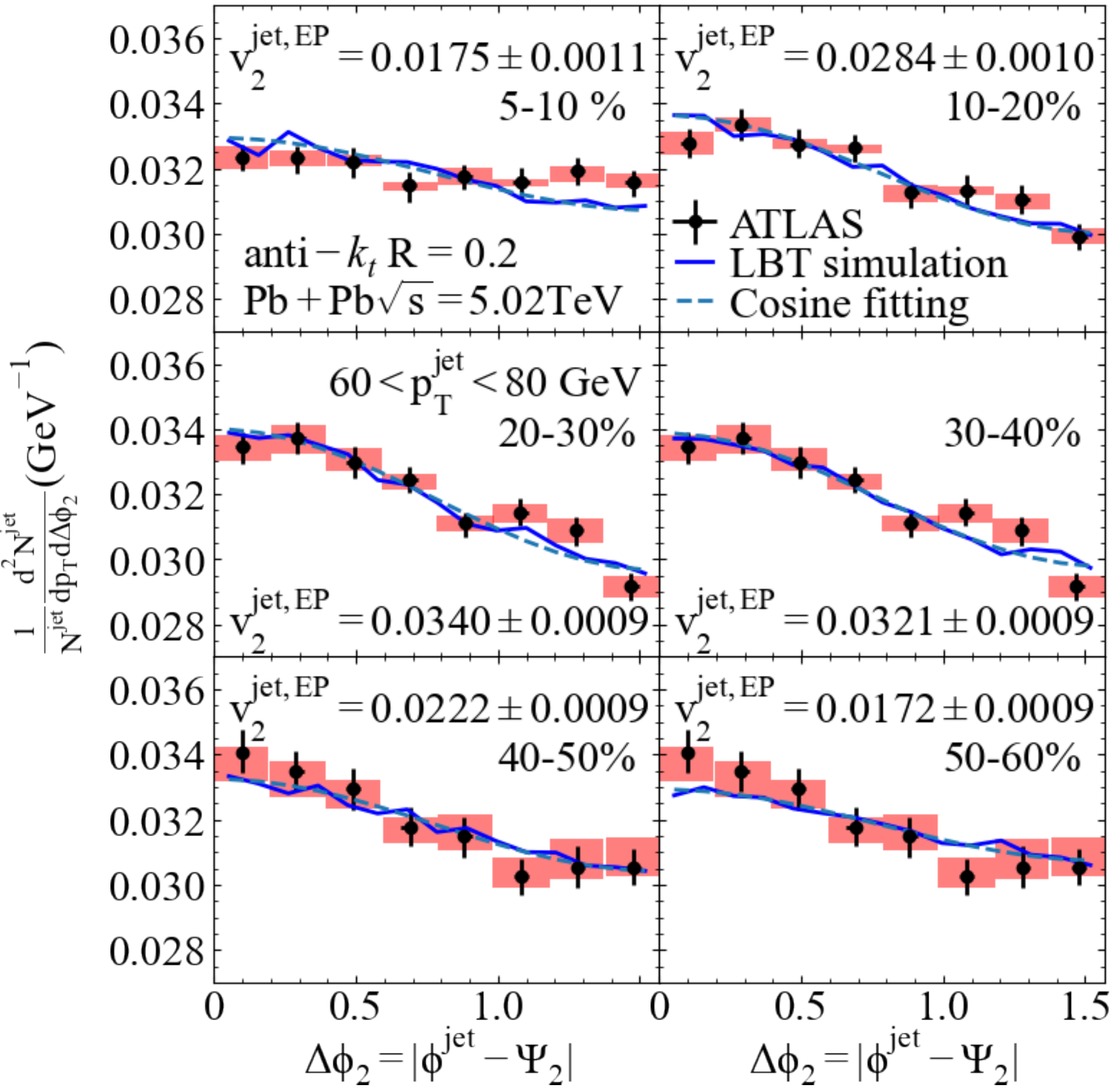}
    \caption{(Color online) Distribution of the difference of the jet azimuthal angle $\phi^\mathrm{jet}$ and the event plane angle $\Psi_{2}$ ($\Delta \phi_{2} = \phi^\mathrm{jet} - \Psi_{2}$) in different centrality bins of Pb+Pb collisions at $\sqrt{s} = 5.02$~TeV. Results calculated from the LBT model (blue solid lines) are compared to the ATLAS data~\cite{ATLAS:2020qxc} (black closed circle) and fitted with a cosine function (blue dash lines).}
    \label{jetphi2_5020}
\end{figure}

Similar results for Pb+Pb collisions at $\sqrt{s} = 5.02$~TeV are presented in Fig.~\ref{jetphi2_5020}. Both the azimuthal angular distributions of the jet yields (cross symbols) from LBT and their cosine function fits (solid lines) are shown for different centrality bins for jets with cone size $R=0.2$ within $60 < p_\mathrm{T} < 80$~GeV/$c$. The extracted $v^\mathrm{jet,EP}_2$ from these cosine fits again first increases and then decreases with the centrality.

\begin{figure}[tbp]
    \includegraphics[width=8cm]{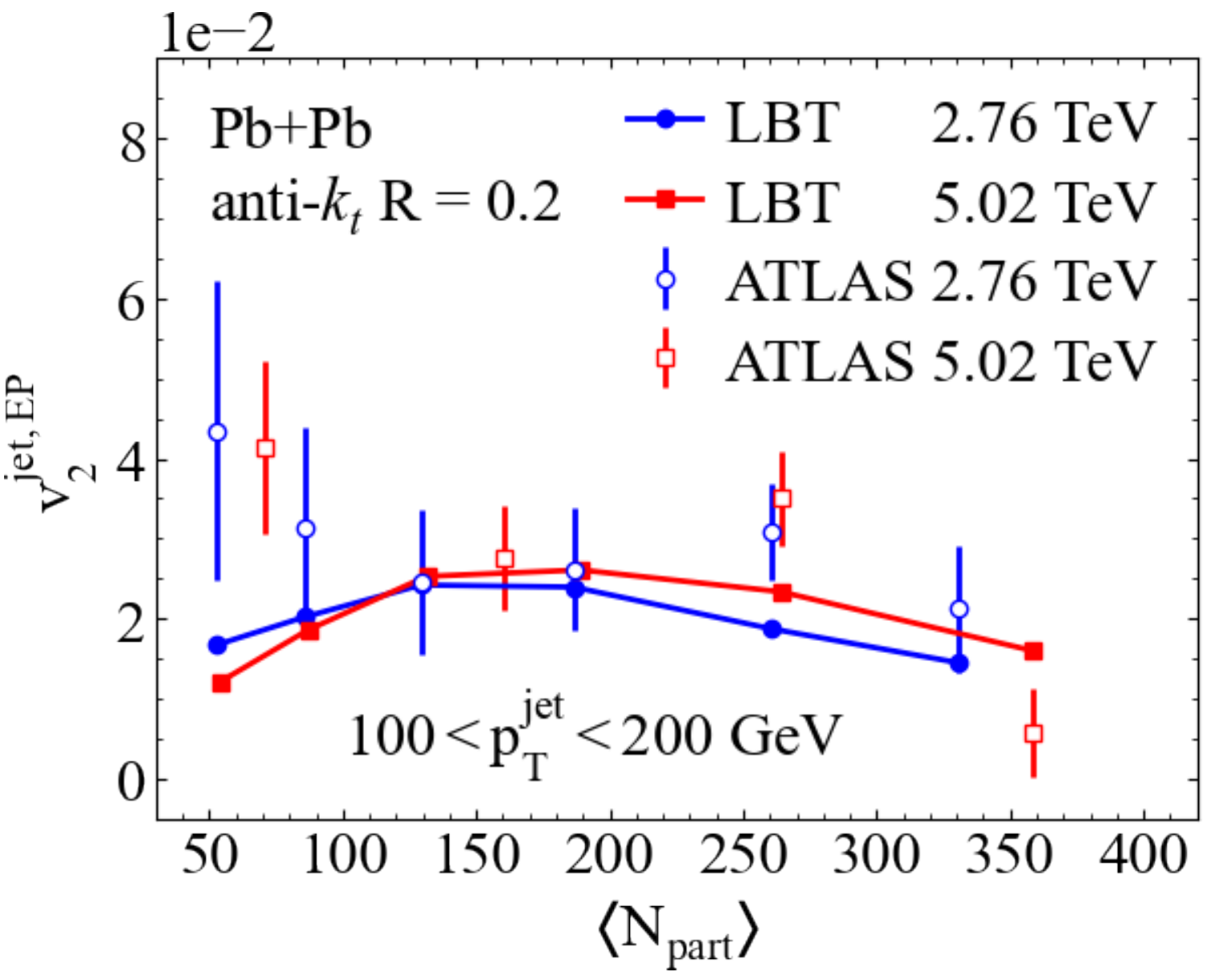}
    \caption{(Color online) Jet elliptic anisotropy $v_{2}^\mathrm{jet, EP}$ in $100 < p_\mathrm{T} < 200$~GeV/$c$ as a function of the number of participant nucleons $N_{\rm part}$ in Pb+Pb collisions at $\sqrt{s} = 2.76$~TeV (blue filled circle) and 5.02~TeV (red filled square) from LBT model calculations as compared to the ATLAS data~\cite{Aad:2013sla,ATLAS:2020qxc} (blue open circle and red open sqaure).}
    \label{v2Npart}
\end{figure}

\begin{figure}[tbp]
    \includegraphics[width=8cm]{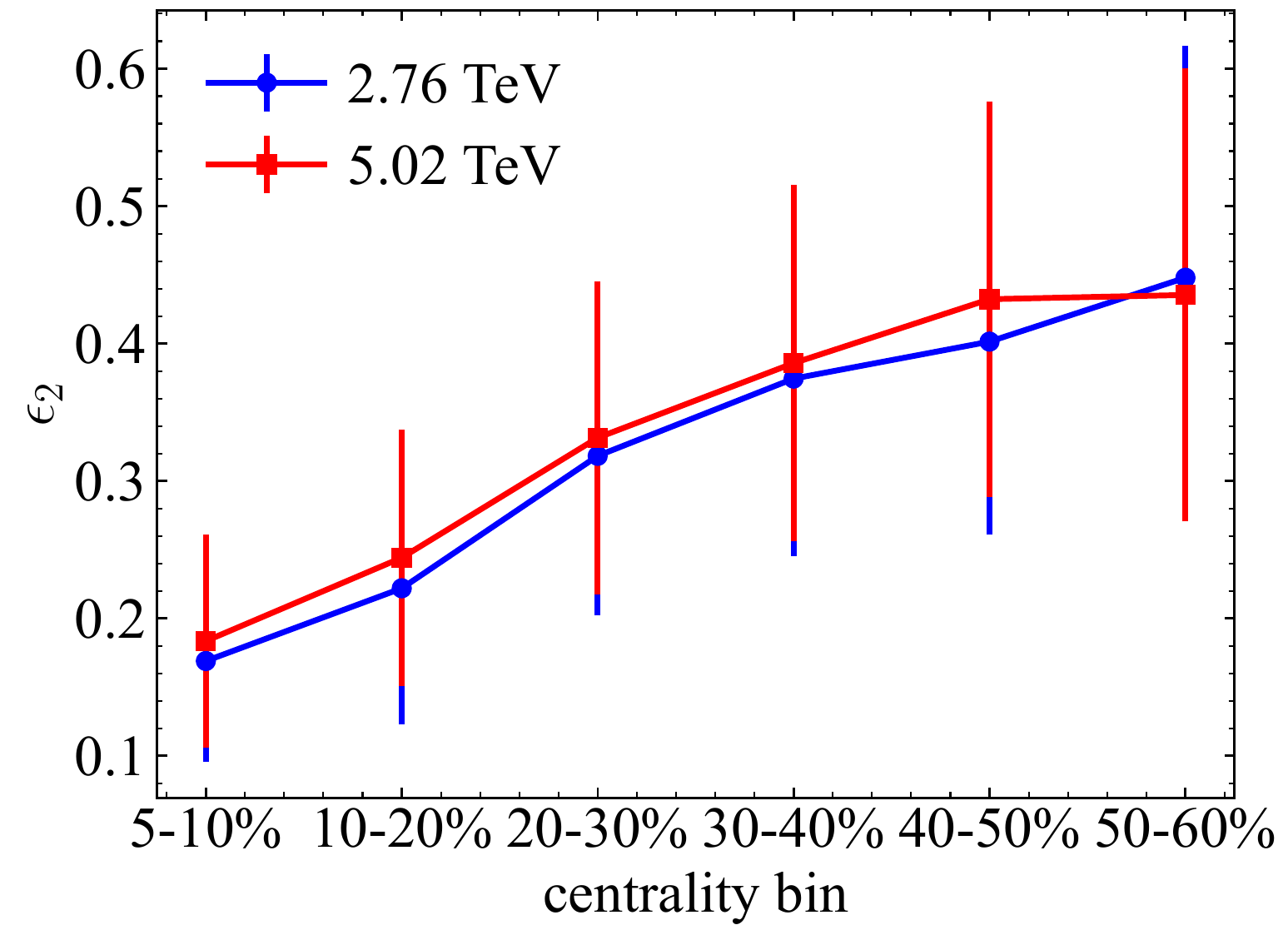}
    \caption{(Color online) The initial second-order eccentricity $\epsilon_2$ as a function of centrality in Pb+Pb collisions at $\sqrt{s} = 2.76$~TeV (blue circle) and 5.02~TeV (red square).}
    \label{ec2}
\end{figure}

In Fig.~\ref{v2Npart}, we summarize $v_{2}^\mathrm{jet, EP}$ for jets with $100 < p_\mathrm{T} < 200 $~GeV/$c$ from analyses of LBT results similar to what are presented in Figs.~\ref{jetphi2_2760} and~\ref{jetphi2_5020} in different centrality bins as a function of the average number of nucleon participants $\langle N_\mathrm{part}\rangle$ in nuclear collisions at both $\sqrt{s}=2.76$ and 5.02 GeV as compared to ATLAS data~\cite{Aad:2013sla,ATLAS:2020qxc}. As we have observed before,  the jet elliptic anisotropy coefficient $v_{2}^\mathrm{jet, EP}$ in these collisions first increases and then decreases with the participant number (as the centrality decreases), due to the competing effects between the geometric anisotropy and the initial size/temperature of the QGP medium. This centrality dependence from LBT simulations are consistent with the ATLAS data for central and semi-central Pb+Pb collisions at the LHC energies. The experimental data at more peripheral collisions are, however, somewhat higher than the LBT results. This might call into question on the validity of complete thermalization assumed in the hydrodynamic model in small systems of peripheral collisions. The trigger bias and the neglect of impact-parameter dependence of nucleon-nucleon collisions in peripheral nuclear collisions \cite{Loizides:2017sqq} can potentially affect jet quenching and anisotropy analyses in these peripheral events. This can be clarified in the future by comparisons with experimental data from central light-ion, for example O+O,  collisions \cite{Huss:2020dwe} or $p$+A collisions~\cite{Xie:2020zdb}.

Though there is no significant difference between experimental data on $v_{2}^\mathrm{jet,EP}$ at the two colliding energies beyond systematic and statistic errors as shown in Fig.~\ref{v2Npart}, there is a small but visible difference between LBT results at these two colliding energies. The small difference also depends on the centrality or the number of participant nucleons.  For large values of $\langle N_\mathrm{part}\rangle $ in central and semi-central collisions, $v_{2}^\mathrm{jet, EP}$ at $\sqrt{s}=5.02$~TeV is slightly larger than at 2.76~TeV. 
This variation can come from a combination of the colliding energy dependence of the jet energy loss (parton density at 5.02~TeV is higher than at 2.76~TeV), initial jet spectra (the spectra at 5.02~TeV is flatter than at 2.76~TeV) and the initial geometric anisotropy. This ordering is also similar to that of the elliptic flow coefficients for soft bulk hadrons in all centrality bins as summarized in Tab.~\ref{table:v2_mean_std}.  However, unlike $v_2^{\rm soft}$, the order for $v_{2}^\mathrm{jet, EP}$ from LBT is reversed for small values of $\langle N_\mathrm{part}\rangle$ in peripheral collisions as seen in Fig.~\ref{v2Npart}. Since $v_{2}^\mathrm{jet,EP}$ is directly driven by the initial geometric anisotropy of the QGP medium $\epsilon_2$ instead of the final momentum anisotropy of soft hadrons $v_2^{\rm soft}$, it is illustrative to exam the colliding energy dependence of the $\epsilon_2$ coefficient from the AMPT model in Fig.~\ref{ec2}. We indeed observe that the average values of $\epsilon_2$ of the QGP medium in Pb+Pb collisions at $\sqrt{s}=5.02$~ TeV are larger than that at 2.76 TeV in central and semi-central collisions, except in the very peripheral collisions where the order is also reserved. A possible cause of the reversed ordering of $v_{2}^\mathrm{jet,EP}$ and  $\epsilon_2$ in the colliding energy in peripheral collisions is the amount of the geometric fluctuation of the corresponding bulk media. The fluctuation is expected to be larger at lower colliding energy, especially in peripheral collisions.


\subsection{$p_\mathrm{T}$ dependence and effect of event fluctuation}

Apart from extracting $v_{n}^\mathrm{jet}$ from the cosine function fit, we also use the following two methods to evaluate the jet anisotropy. The first one is equivalent to the cosine function fit, i.e., the event plane (``EP'') method, with the definition of the anisotropy as,
\begin{equation}
    v_{n}^\mathrm{jet, EP} = \langle \langle \cos(n[\phi^\mathrm{jet} - \Psi_{n}]) \rangle \rangle,
    \label{vnjet_noFluc0}
\end{equation}
where the symbol $\langle\langle\ldots\rangle\rangle$ denotes the average over events. This event plane method does not take into account the fluctuation of the bulk $v_n^{\rm soft}$ within a given centrality bin. In the second approach, the scalar product (``SP") method takes into account the correlation with $v_n^{\rm soft}$ fluctuation in the jet anisotropy,
\begin{equation}
    v_{n}^\mathrm{jet, SP} = \frac{\langle \langle v_{n}^{\rm soft} \cos(n[\phi^\mathrm{jet} - \Psi_{n}]) \rangle \rangle}{\sqrt{\langle {v_{n}^{\rm soft}}^{2} \rangle}},
    \label{vnjet_Fluc0}
\end{equation}
where $v_n^{\rm soft}$ represents the anisotropy of the soft bulk hadrons in one event, while $\sqrt{\langle {v_n^{\rm soft}}^2\rangle}$ denotes the root-mean-square average of $v_n^{\rm soft}$ within a given centrality bin. 

\begin{figure}[tb]
    \includegraphics[width=8cm]{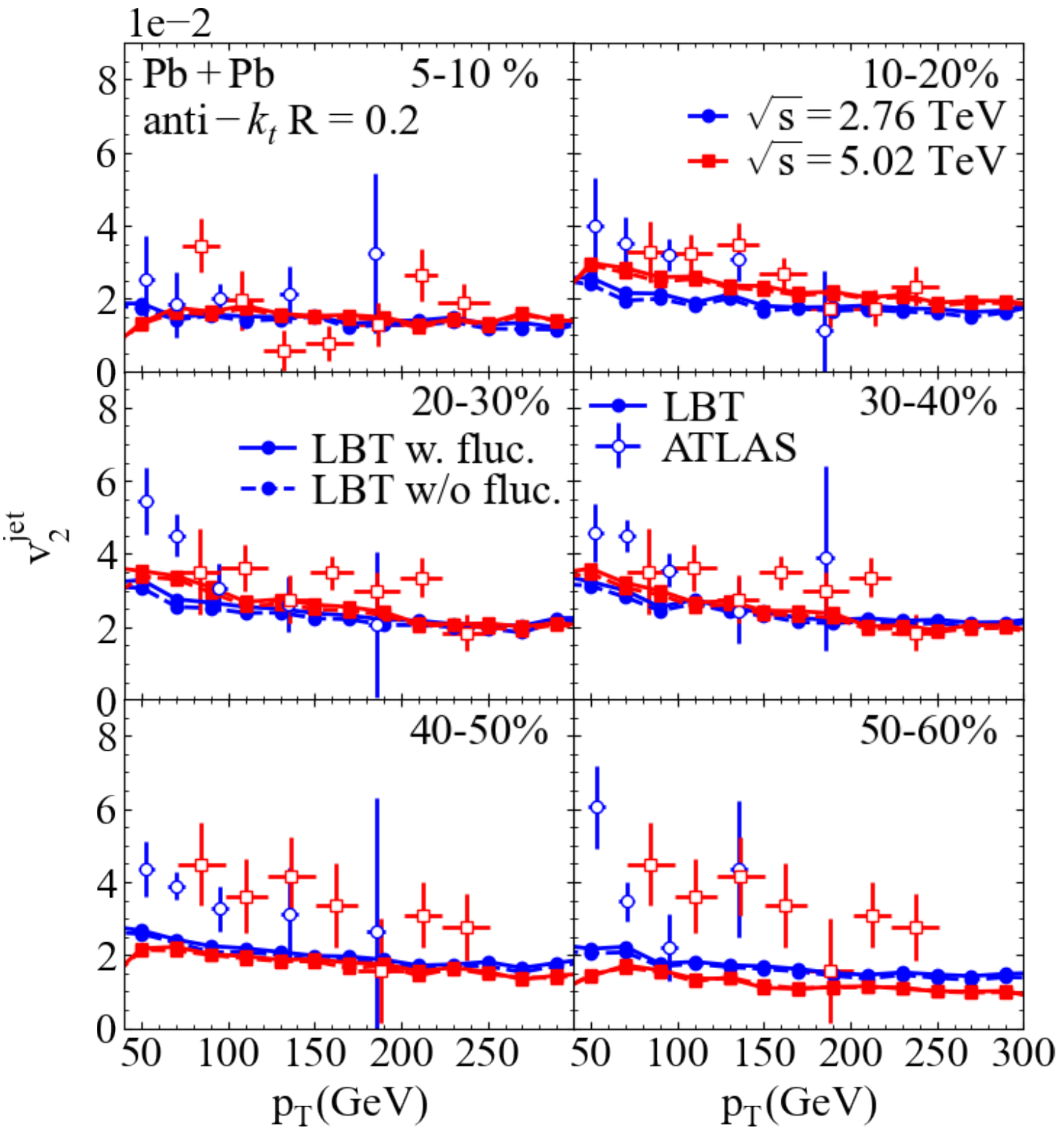}
    \caption{(Color online) Jet elliptic anisotropy coefficient $v_{2}^\mathrm{jet}$ as a function of the jet $p_\mathrm{T}$ from the LBT model (closed marker) as compared to the ATLAS data~\cite{Aad:2013sla,ATLAS:2020qxc} (open marker) in different centrality ranges of Pb+Pb collisions at $\sqrt{s} = 2.76$~TeV (blue circle) and $5.02$~TeV (red square ). $v_{2}^\mathrm{jet, SP}$ is labeled as ``with (soft $v_{2}$) fluctuations" (solid lines), and $v_{2}^\mathrm{jet, EP}$ is labeled as ``without fluctuations" (dashed lines).}
    \label{jetv2_twoEnergy}
\end{figure}

 In Fig.~\ref{jetv2_twoEnergy} we show the $p_\mathrm{T}$ dependence of $v_{2}^\mathrm{jet}$ with the above two different methods for Pb+Pb collisions with different centralities at both $\sqrt{s}=2.76$ and 5.02~TeV, compared to the ATLAS data. The LBT results with the event plane (scalar product) method are shown as solid (dashed) lines that takes (does not take) into account of the fluctuation of soft hadron anisotropies $v_n^{\rm soft}$. We find that the effect of the event-by-event fluctuation of the soft hadron $v_2^{\rm soft}$ on $v_{2}^\mathrm{jet}$ is negligibly small.  We observe that $v_2^{\rm jet}$ in all centrality classes has a weak transverse momentum dependence, decreasing slightly with $p_{\rm T}$. This is consistent with the $p_{\rm T}$ dependence of the single inclusive jet suppression factor $R_{\rm AA}(p_{\rm T})$ which increases slightly with $p_{\rm T}$ as a result of the interplay between the $p_{\rm T}$ dependence of the jet energy loss and the shape of the initial jet spectra~\cite{He:2018xjv}. Again, the LBT model can describe the ATLAS data well except in very peripheral Pb+Pb collisions.

\begin{figure}[tbp]
    \includegraphics[width=8cm]{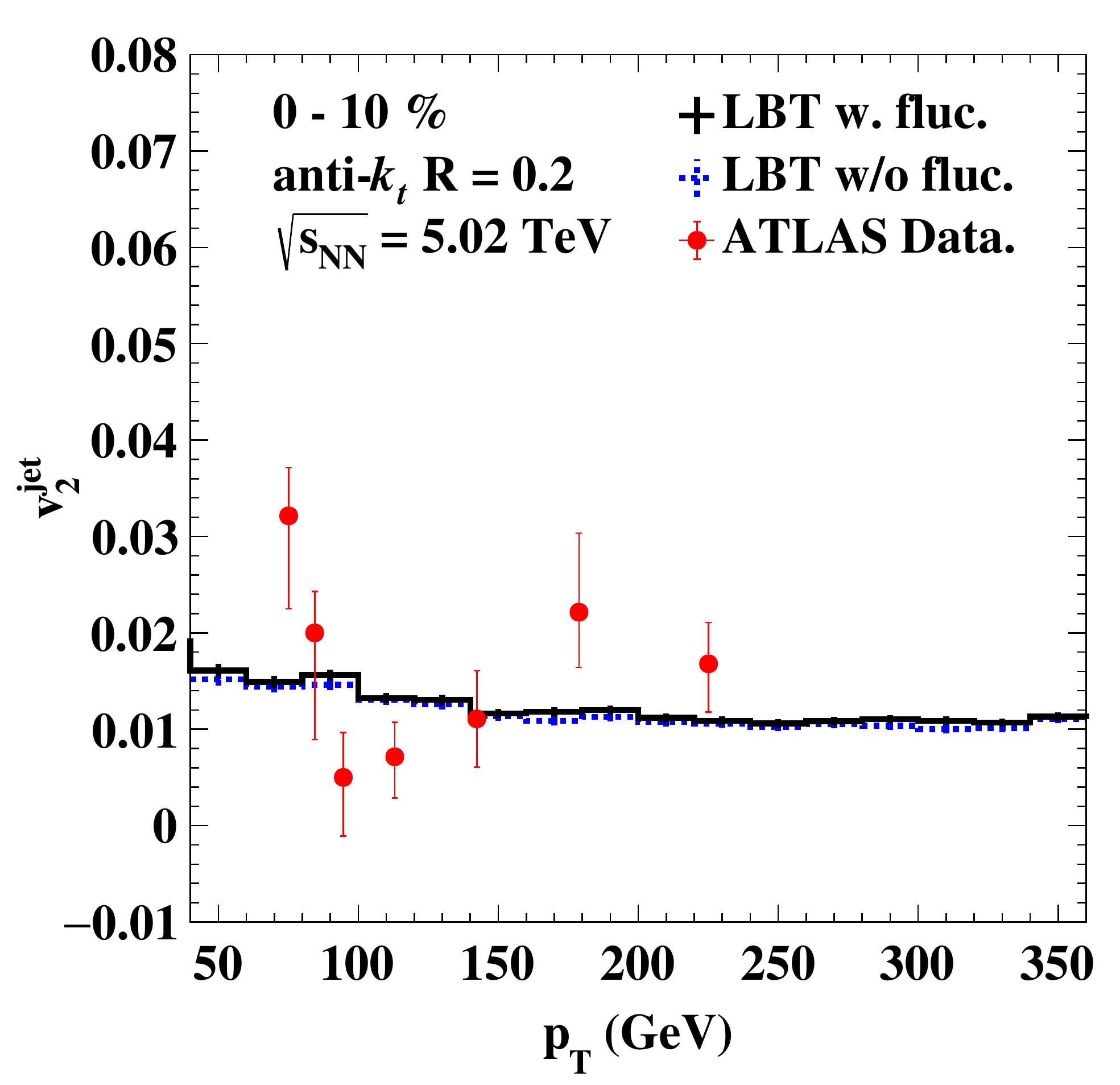}
    \caption{(Color online) Jet elliptic anisotropy coefficient $v_{2}^\mathrm{jet}$ as a function of the jet $p_\mathrm{T}$ from the LBT model calculation (lines) as compared to the ATLAS data~\cite{ATLAS:2020qxc} (markers) in 0-10\% Pb+Pb collisions at $\sqrt{s} = 5.02$~TeV. $v_{2}^\mathrm{jet, SP}$ is labeled as ``with (soft $v_{2}$) fluctuations" (solid lines), and $v_{2}^\mathrm{jet, EP}$ is labeled as ``without fluctuations" (dashed lines).}
    \label{jetv2_new}
\end{figure}

For a closer look at the effect of soft hadron $v_2^{\rm soft}$ fluctuation on jet $v_2^{\rm jet}$, we present in Fig.~\ref{jetv2_new} a comparison between the LBT results and the recent ATLAS data~\cite{ATLAS:2020qxc} on $v_{2}^\mathrm{jet}$ in Pb+Pb collisions at 5.02~TeV. Our LBT calculation provides a reasonable description of the experimental data.
The inclusion of the soft hadron  $v_2^{\rm soft}$ fluctuation increases the LBT results slightly on $v_{2}^\mathrm{jet}$, though the effect is extremely small. This small relative difference is consistent with the estimate $(\delta v_2^{\rm soft}/v_2^{\rm soft})^2\sim 0.01-0.02$ according to Tab.~\ref{table:v2_mean_std}. It is also consistent with the findings for the elliptic flow of single inclusive hadrons in an earlier study~\cite{Cao:2017umt}. This is, however, in sharp contrast to the conclusion in Ref.\cite{Noronha-Hostler:2016eow} where the effect of bulk flow fluctuation was found significantly larger for high $p_\mathrm{T}$ single inclusive hadrons.


Again, the experimental data in Fig.~\ref{jetv2_twoEnergy} show no significant dependence on the colliding energy. The LBT results have the same small colliding energy dependence, consistent with the findings in Fig.~\ref{v2Npart}. The jet elliptic anisotropy $v_{2}^\mathrm{jet}$ is slightly larger or smaller in collisions at $\sqrt{s} = 2.76$~TeV than at 5.02~TeV  depending on the centrality, due to the centrality dependence of the different bulk geometric anisotropy $\epsilon_2$ produced at these two colliding energies as we have just discussed in the above subsection.


\begin{figure}[tbp]
    \includegraphics[width=8cm]{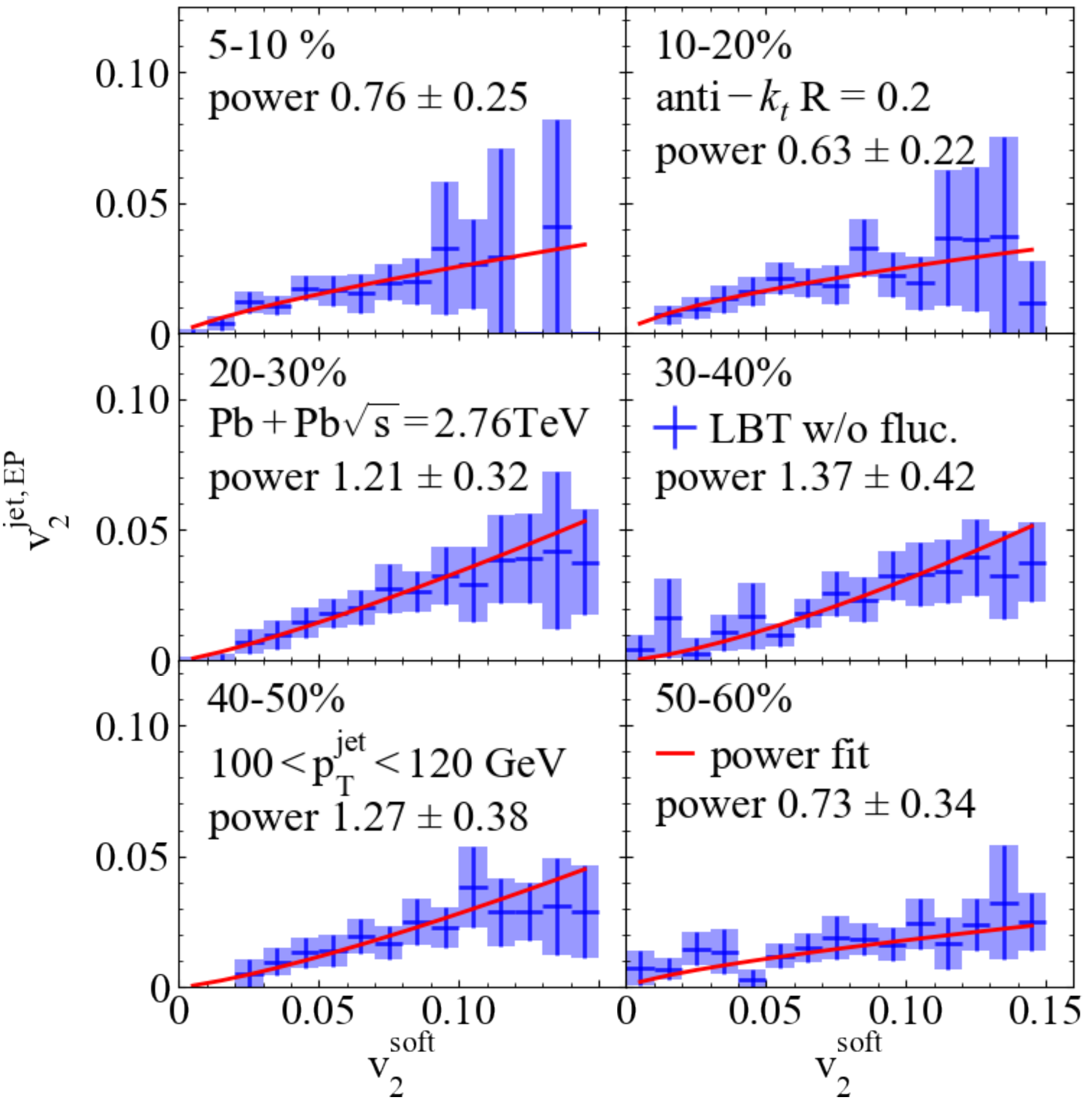}
    \caption{(Color online) Correlations between $v_{2}^\mathrm{jet, EP}$ and the bulk $v_2^\mathrm{soft}$ in Pb+Pb collisions at $\sqrt{s} = 2.76$~TeV with jet transverse momentum $100 < p_\mathrm{T} < 120$~GeV/$c$. The LBT results (blue points with error boxes) is fitted with a power law correlation (red lines) in each centrality bin with the power index as indicated.}
    \label{v2jetv20_100_2760}
\end{figure}

\begin{figure}[tbp]
    \includegraphics[width=8cm]{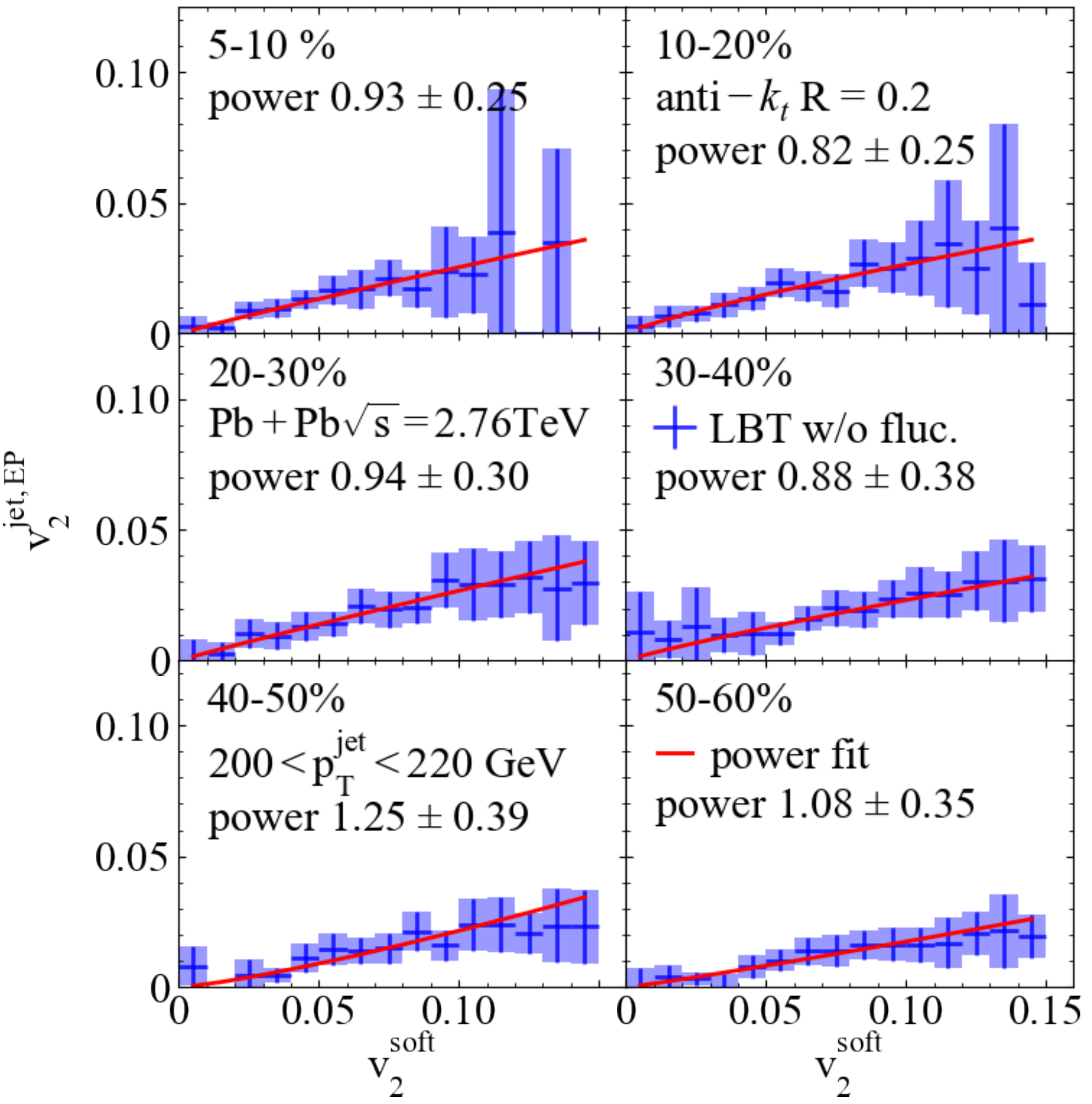}
    \caption{(Color online) Correlations between $v_{2}^\mathrm{jet, EP}$ and the bulk $v_2^\mathrm{soft}$ in Pb+Pb collisions at $\sqrt{s} = 2.76$~TeV with jet transverse momentum $200 < p_\mathrm{T} < 220$~GeV/$c$. The LBT results (blue points with error boxes) is fitted with a power law correlation (red lines) in each centrality bin with the power index as indicated.}
    \label{v2jetv20_200_2760}
\end{figure}

\begin{figure}[tbp]
    \includegraphics[width=8cm]{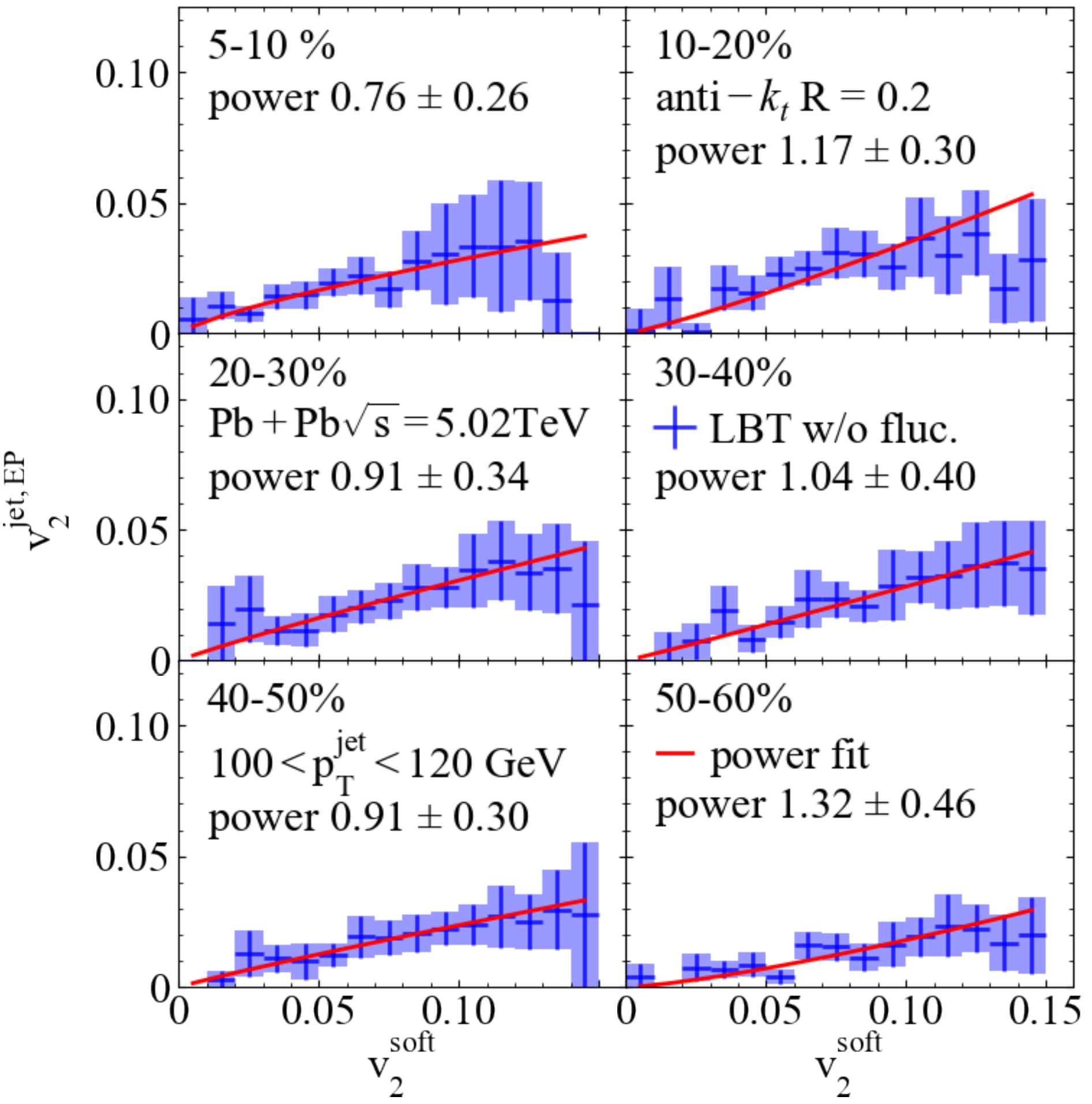}
    \caption{(Color online) Correlations between $v_{2}^\mathrm{jet, EP}$ and the bulk $v_2^\mathrm{soft}$ in Pb+Pb collisions at $\sqrt{s} = 5.02$~TeV with jet transverse momentum $100 < p_\mathrm{T} < 120$~GeV/$c$. The LBT results (blue points with error boxes) is fitted with a power law correlation (red lines) in each centrality bin with the power index as indicated.}
    \label{v2jetv20_100_5020}
\end{figure}

\begin{figure}[tbp]
    \includegraphics[width=8cm]{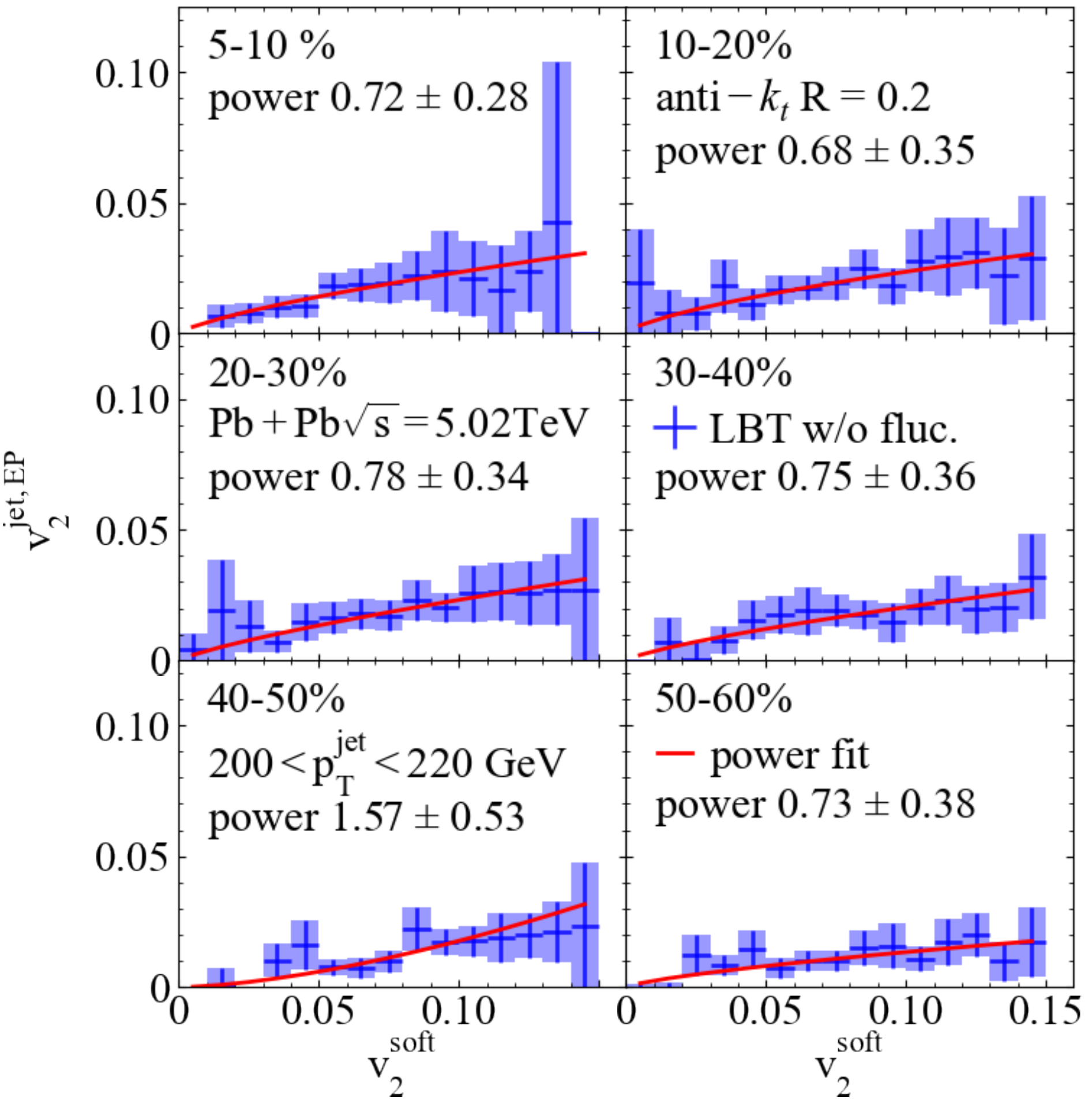}
    \caption{(Color online) Correlations between $v_{2}^\mathrm{jet, EP}$ and the bulk $v_2^\mathrm{soft}$ for different centralities of $\sqrt{s} = 5.02$~TeV Pb+Pb collisions with jet transverse momentum within $200 < p_\mathrm{T} < 220$~GeV. The LBT results (blue points with error boxes) is fitted with a power law (red lines).}
    \label{v2jetv20_200_5020}
\end{figure}

\subsection{Soft and hard correlation}

The geometric anisotropy of the QGP medium is expected to be the dominant factor that determines the jet anisotropy $v_{n}^\mathrm{jet}$ due to the length dependence of transverse momentum broadening and parton energy loss. This geometric anisotropy fluctuates from event to event even for a given centrality class as reflected by the fluctuation of the anisotropic flow coefficients of soft bulk hadrons shown in Figs.~\ref{v2soft} and \ref{v3soft}. Since the anisotropic flows of soft hadrons are also driven by the initial geometric anisotropies of the QGP medium, one should expect a direct event-by-event correlation between the anisotropic flow of soft hadrons $v_2^\mathrm{soft}$ and the jet anisotropy $v_2^{\rm jet}$.

In Figs.~\ref{v2jetv20_100_2760}, \ref{v2jetv20_200_2760}, \ref{v2jetv20_100_5020} and~\ref{v2jetv20_200_5020}, we show the correlation between the soft hadron $v_2^\mathrm{soft}$ and the jet elliptic anisotropy $v_{2}^\mathrm{jet}$ in different jet $p_\mathrm{T}$ ranges for different centrality classes of Pb+Pb collisions at the LHC energies as simulated by LBT and CLVisc. Results for both $\sqrt{s}=2.76$~TeV (Figs.~\ref{v2jetv20_100_2760} and~\ref{v2jetv20_200_2760}) and 5.02~TeV (Figs.~\ref{v2jetv20_100_5020} and~\ref{v2jetv20_200_5020}) are presented. In each panel of these figures, blue crosses represent results from our LBT and CLVisc calculations, while red curves represent a power law fit in the following form,
\begin{equation}
    v_{2}^\mathrm{jet} = \alpha  {(v_{2}^\mathrm{soft})}^{\beta},
    \label{power_fit}
\end{equation}
with $\alpha$ as an overall normalization factor and $\beta$ as the power index. Both anisotropies are analyzed with event plane method. Note that for any given value of $v_2^{\rm soft}$ in each hydro event, the jet anisotropy $v_2^{\rm jet}$ from the event plane method in Eq.~(\ref{vnjet_noFluc0}) and the scalar product method in Eq.~(\ref{vnjet_Fluc0}) coincide. One observes that the power indices $\beta$ from the fitting are all around 1, indicating an approximately linear correlation between the jet $v_2^\mathrm{jet,EP}$ and the bulk $v_2^\mathrm{soft}$ due to the fluctuation of the bulk medium.  This linear correlation is quite interesting and not necessarily straightforward since the bulk hadron anisotropy arises from collective expansion while the jet anisotropy is caused by the length dependence of parton energy loss. The initial geometrical anisotropy of the expanding QGP fireball is the link underlying this correlation. We have tried to include only elastic processes in the LBT simulations with increased effective coupling constant so that one can still fit the experimental data on single inclusive jet suppression $R_{\rm AA}^{\rm jet}$. This model simulation leads to only a slightly different length dependence of jet energy loss. However the resultant soft and hard correlation between $v_2^\mathrm{jet}$ and $v_2^\mathrm{soft}$ still remains approximately linear. One probably needs to have a joint analysis of the centrality dependence of the single inclusive jet suppression $R_{\rm AA}^{\rm jet}$ and jet anisotropy $v_2^\mathrm{jet}$ and the soft hard correlation between $v_2^\mathrm{jet}$ and $v_2^\mathrm{soft}$ in order to provide a stringent constraint on the jet transport dynamics. This is however beyond the scope of the study in this paper.

\section{Single inclusive jet anisotropy $v^\mathrm{jet}_3$}
\label{sec:v3}

In addition to the elliptic flow coefficient, we can also study the triangular flow coefficient $v^\mathrm{jet}_3$ of jets, which helps place more stringent constraints on the jet transport dynamics  inside the QGP medium as well as the effect of bulk medium fluctuation on jet observables.

\begin{figure}[t]
    \includegraphics[width=8cm]{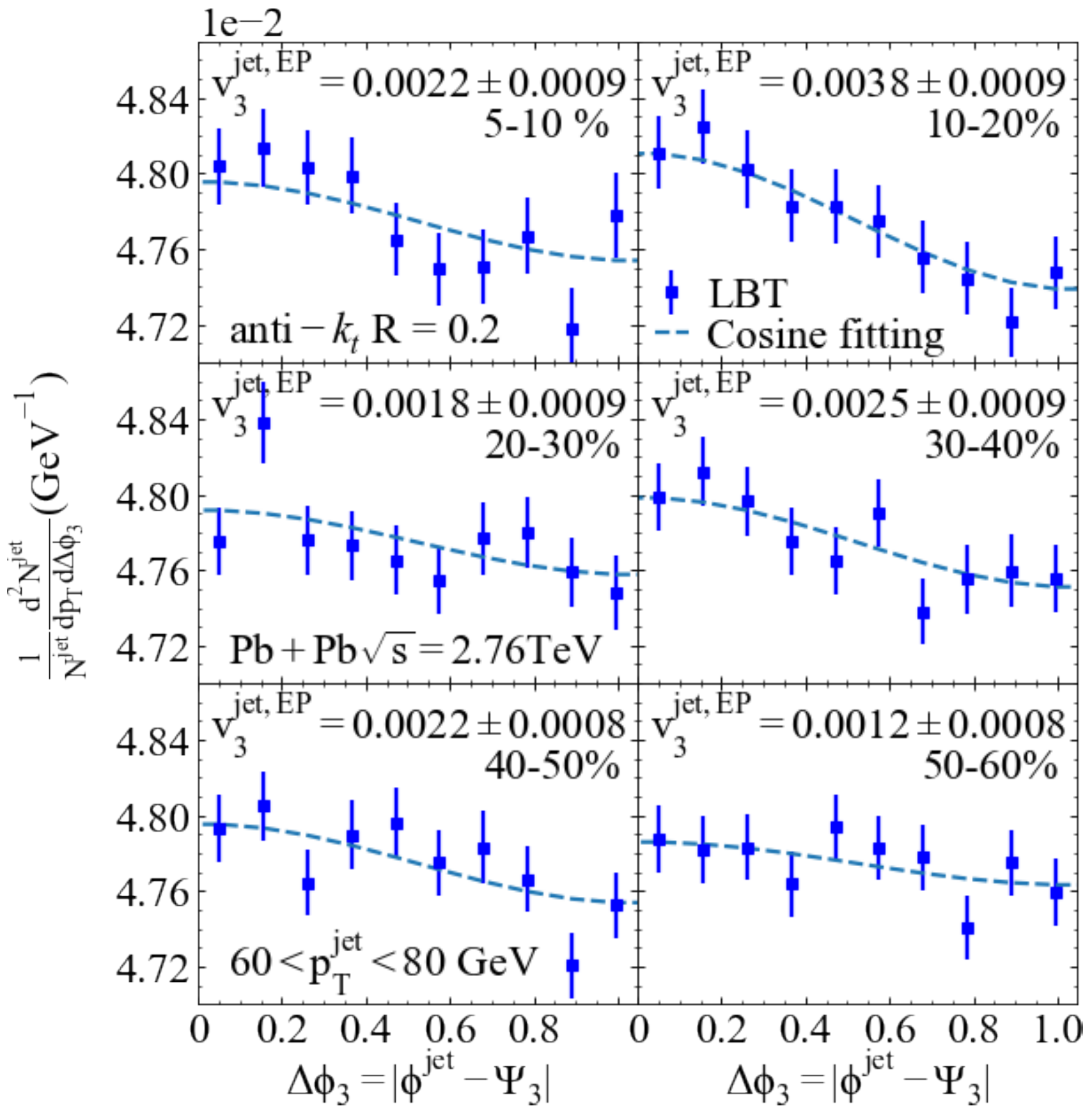}
    \caption{(Color online) Distributions of the difference of the jet azimuthal angle $\phi^\mathrm{jet}$ and the event plane angle $\Psi_{3}$ ($\Delta \phi_{3} = \phi^\mathrm{jet} - \Psi_{3}$) for different centrality bins of Pb+Pb collisions at $\sqrt{s} = 2.76$~TeV. Results calculated from the LBT model (blue solid squares) are fitted with a cosine function (blue dash lines).}
    \label{jetphi3_2760}
\end{figure}

\begin{figure}[tbp]
    \includegraphics[width=8cm]{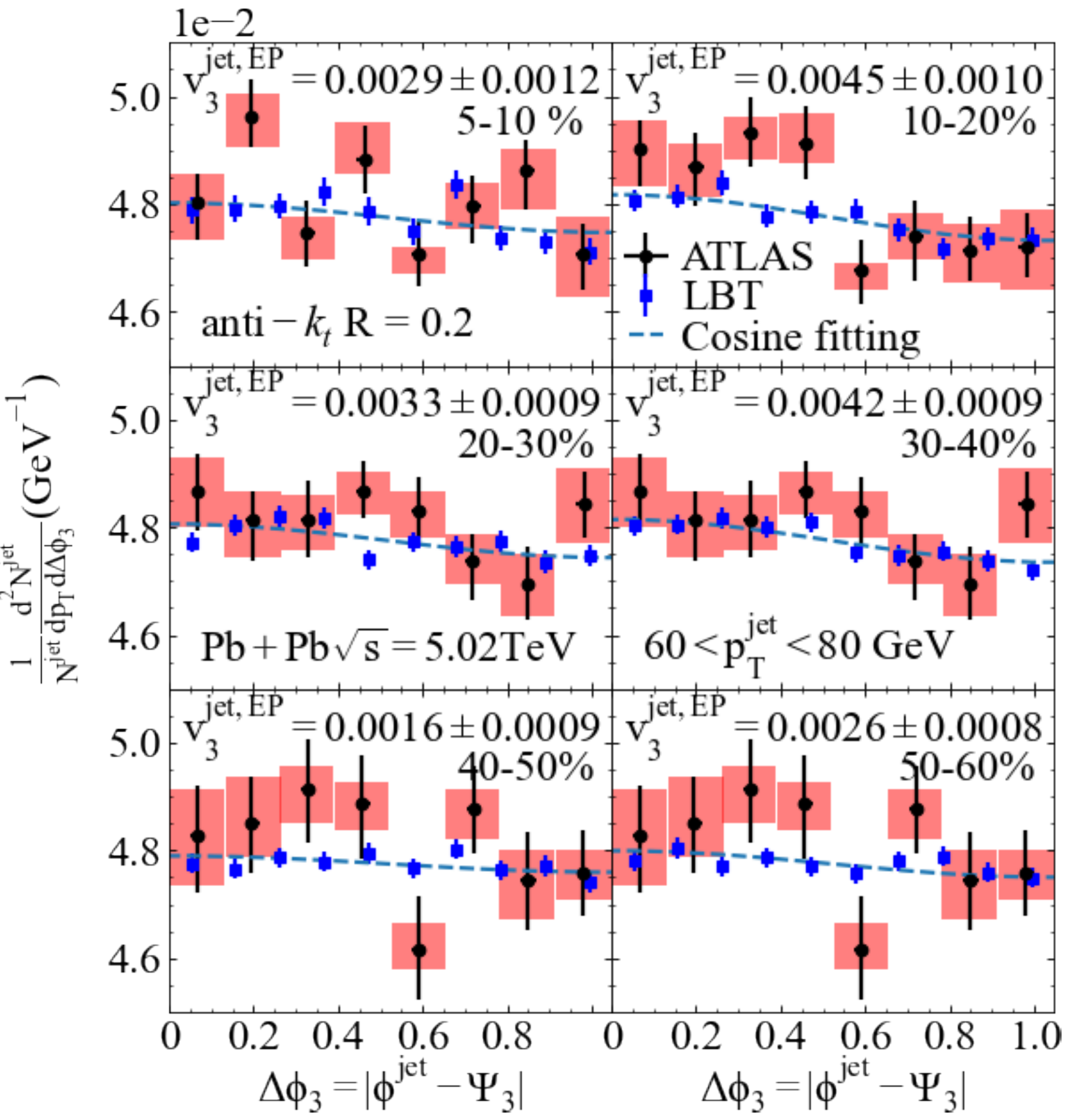}
    \caption{(Color online) Distributions of the difference of the jet azimuthal angle $\phi^\mathrm{jet}$ and the event plane angle $\Psi_{3}$ ($\Delta \phi_{3} = \phi^\mathrm{jet} - \Psi_{3}$) for different centrality bins of Pb+Pb collisions at $\sqrt{s} = 5.02$~TeV. Results calculated from the LBT model (blue solid squares) are compared with the ATLAS data~\cite{ATLAS:2020qxc} (black closed circle) and fitted with a cosine function (blue dash lines).}
    \label{jetphi3_5020}
\end{figure}

Similar to the discussion about the elliptic jet anisotropy in the previous section, we first present the azimuthal angle distribution of single inclusive jets within $60 < p_\mathrm{T} < 80$~GeV in Pb+Pb collisions at $\sqrt{s}=2.76$~TeV in Fig.~\ref{jetphi3_2760}, and at 5.02~TeV in Fig.~\ref{jetphi3_5020}. The horizontal axis $\Delta \phi_3\equiv \phi^\mathrm{jet}-\Psi_{3}$ is defined as the difference between the azimuthal angle of the jet $\phi^\mathrm{jet}$ and the $3^\mathrm{rd}$-order event plane angle $\Psi_{3}$ of the bulk hadrons. We then fit the jet distribution in $\Delta \phi_3$ from the LBT calculations (solid blue squares) with a cosine function $3 / (\pi \Delta p_\mathrm{T}) ( 1 + 2 v_{3}^\mathrm{jet,EP} \cos(3 \Delta \phi_{3}))$ (dashed lines), where the $3/\pi$ factor is introduced to normalize the distribution function within $[0,\pi/3]$. Since such a fitting corresponds to the event plane method, we label the jet triangular anisotropy coefficient as $v_3^{\rm jet,EP}$. The extracted values of $v_{3}^\mathrm{jet,EP}$ are presented in different panels for various classes of centrality.

\begin{figure}[tbp]
    \includegraphics[width=8cm]{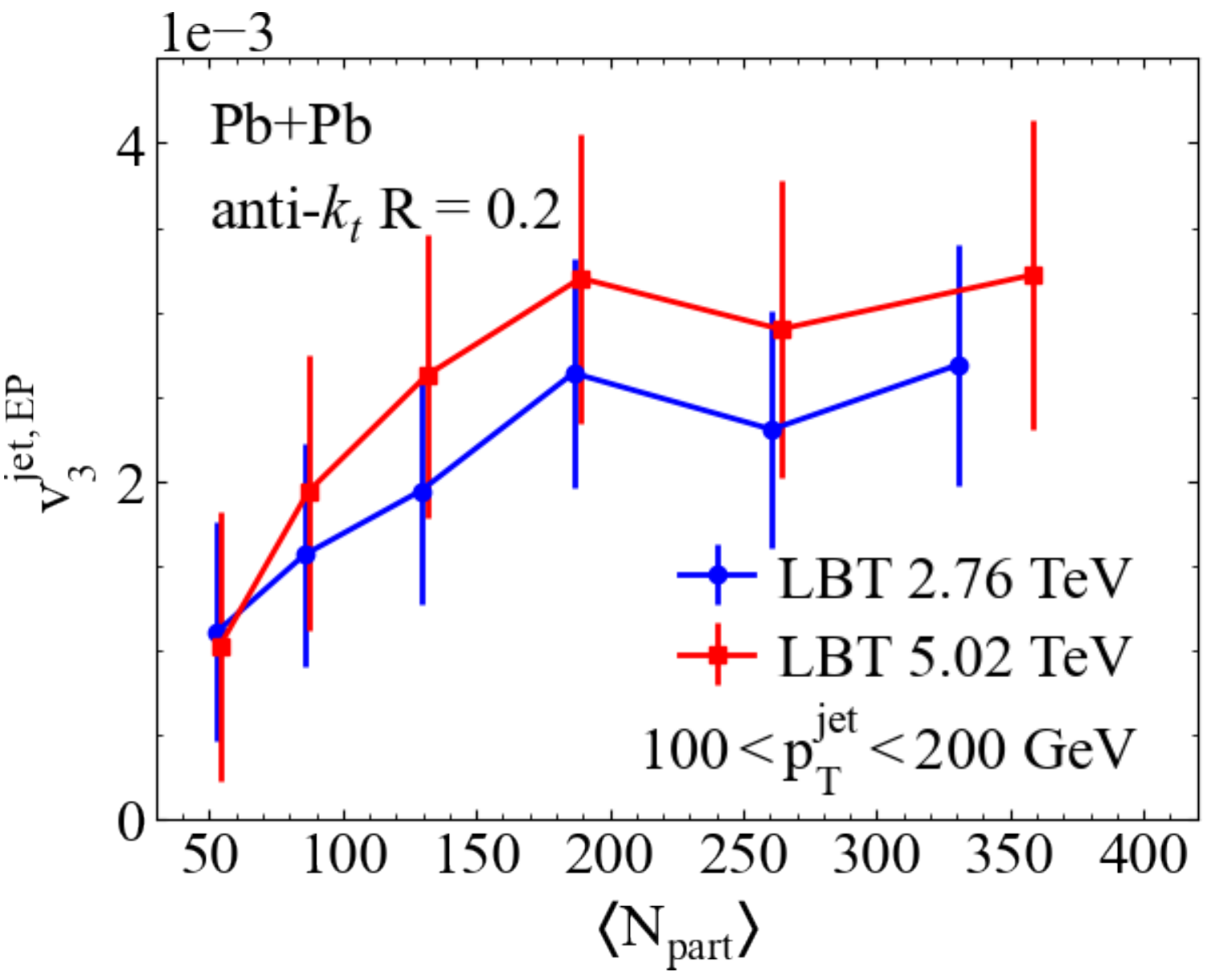}
    \caption{(Color online) The jet $v_{3}^\mathrm{jet, EP}$ within $100 < p_\mathrm{T} < 200$~GeV/$c$ as a function of the number of participant nucleons $N_{\rm part}$ in Pb+Pb collisions at $\sqrt{s} = 2.76$~TeV (blue circles with line) and 5.02~TeV (red squares with line).}
    \label{v3Npart}
\end{figure}

To summarize the centrality dependence of $v^\mathrm{jet}_3$ and compare the results at two colliding energies, we present $v^\mathrm{jet,EP}_3$ as a function of the number of participant nucleons in Fig.~\ref{v3Npart} for jets with the cone size $R=0.2$ and transverse momentum  $100<p_\mathrm{T}<200$~GeV/$c$ in different centrality classes of Pb+Pb collisions at $\sqrt{s}=2.76$ and 5.02~TeV. One observes that $v^\mathrm{jet,EP}_3$ in general is larger in central collisions (large $N_{\rm part}$) than in peripheral collisions (small $N_{\rm part}$). This is different from the centrality dependence of the elliptic jet anisotropy coefficient $v^\mathrm{jet}_2$ as previously shown in Fig.~\ref{v2Npart}.
The triangular geometric anisotropy $\epsilon_3$ of the bulk medium results from the initial state fluctuation while the elliptic geometric anisotropy $\epsilon_2$ is mainly caused by the shape of the nuclear overlap in non-central collisions, though $\epsilon_2$ in the most central collisions also comes from initial fluctuations. As illustrated in Tab.~\ref{table:v3_mean_std}, the bulk $v_3^{\rm soft}$ has a very weak centrality dependence. This should be the same for the triangular geometric anisotropy of the bulk medium. Therefore, larger energy loss of jets in more central collisions naturally gives rise to larger $v^\mathrm{jet,EP}_3$ than that in peripheral collisions.  On the other hand, the elliptic geometric anisotropy of the nuclear overlap is strongly correlated with the centrality, decreasing toward central collisions. This decrease can overcome the increased jet energy loss such that the final jet elliptic anisotropy  $v^\mathrm{jet,EP}_2$ also decreases toward central collisions as seen in Fig.~\ref{v2Npart}.

Compared to Fig.~\ref{v2Npart} for the centrality dependence of the jet elliptic anisotropy, we observe that the values of the jet triangular anisotropy $v^\mathrm{jet}_3$ from LBT in Fig.~\ref{v3Npart} are almost an order of magnitude smaller than those of the jet elliptic anisotropy  $v^\mathrm{jet}_2$. However, these are still significantly larger than the triangular anisotropy of large transverse momentum single inclusive hadron spectra predicted in Ref.~\cite{Noronha-Hostler:2016eow}.
We also observe that the jet triangular anisotropy $v^\mathrm{jet,EP}_3$ in Pb+Pb collisions at $\sqrt{s}= 5.02$~TeV from LBT simulations is larger than that at 2.76~TeV for all the centrality classes. Since the geometrical triangular anisotropies of the systems are similar between the two colliding energies as indicated by the bulk triangular flow coefficients shown in Tab.~\ref{table:v3_mean_std}, this change of jet triangular anisotropy with the colliding energy from 2.76 to 5.02 TeV is mainly caused by the increase of jet energy loss as the initial parton density and the jet transport coefficient increase by about 20\%, indicated by the similar increase in the charged hadron multiplicity in the central rapidity region \cite{ALICE:2013jfw,ALICE:2016fbt}.

\begin{figure}[t]
    \includegraphics[width=8cm]{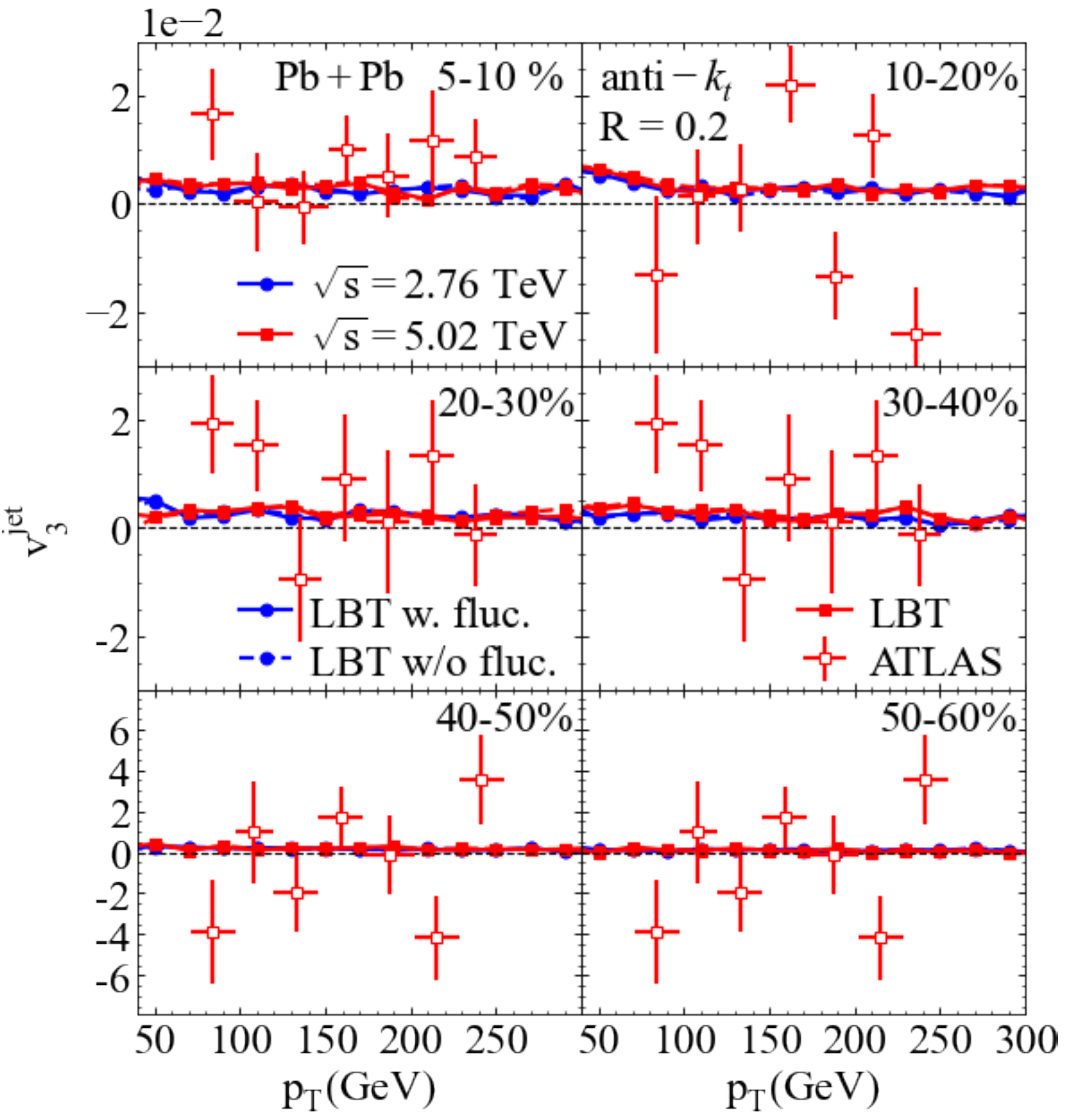}
    \caption{(Color online) The jet $v_{3}^\mathrm{jet}$ as a function of the jet $p_\mathrm{T}$ in different centralities of Pb+Pb collisions at $\sqrt{s} = 2.76$~TeV (blue circles) and $5.02$~TeV (red squares) from the LBT model calculation (closed marker) as compared to the ATLAS data~\cite{ATLAS:2020qxc} (open marker), $v_{3}^\mathrm{jet, SP}$ is labeled as ``with (soft $v_{3}$) fluctuations" (solid lines) while $v_{3}^\mathrm{jet, EP}$ is labeled as ``without fluctuations" (dashed lines).}
    \label{jetv3_twoEnergy_withData}
\end{figure}

\begin{figure}[tbp]
    \includegraphics[width=8cm]{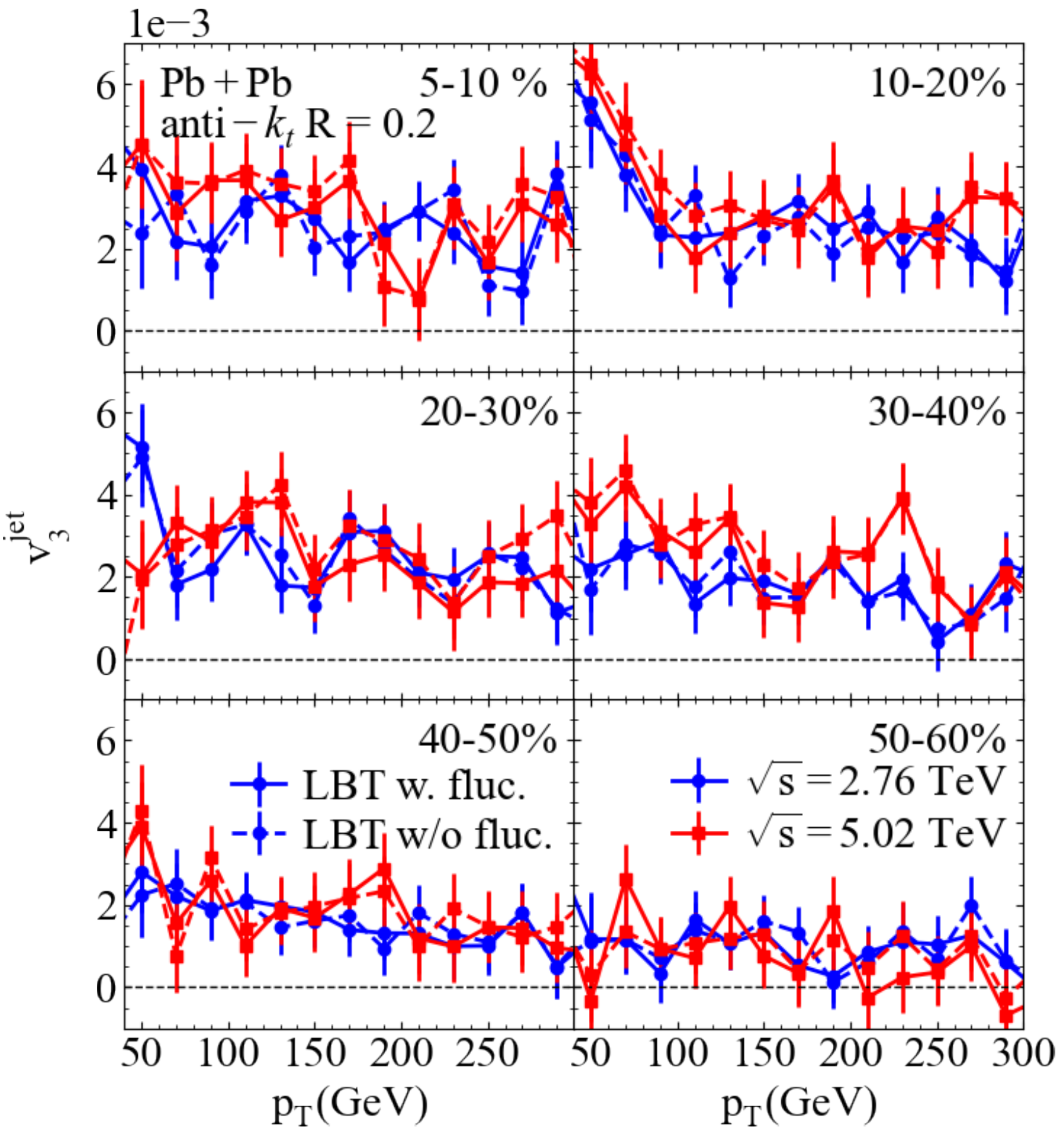}
    \caption{(Color online) The jet $v_{3}^\mathrm{jet}$ as a function of the jet $p_\mathrm{T}$ in different centralities of Pb+Pb collisions at $\sqrt{s} = 2.76$~TeV (blue circles) and $5.02$~TeV (red squares) from the LBT calculations using the event plane analysis $v_{3}^\mathrm{jet, SP}$ (labeled as ``with soft $v_{3}$ fluctuations" -- solid lines) and scalar product analysis $v_{3}^\mathrm{jet, EP}$ (labeled as ``without fluctuations" -- dashed lines).}
    \label{jetv3_twoEnergy}
\end{figure}

\begin{figure}[tbp]
    \includegraphics[width=8cm]{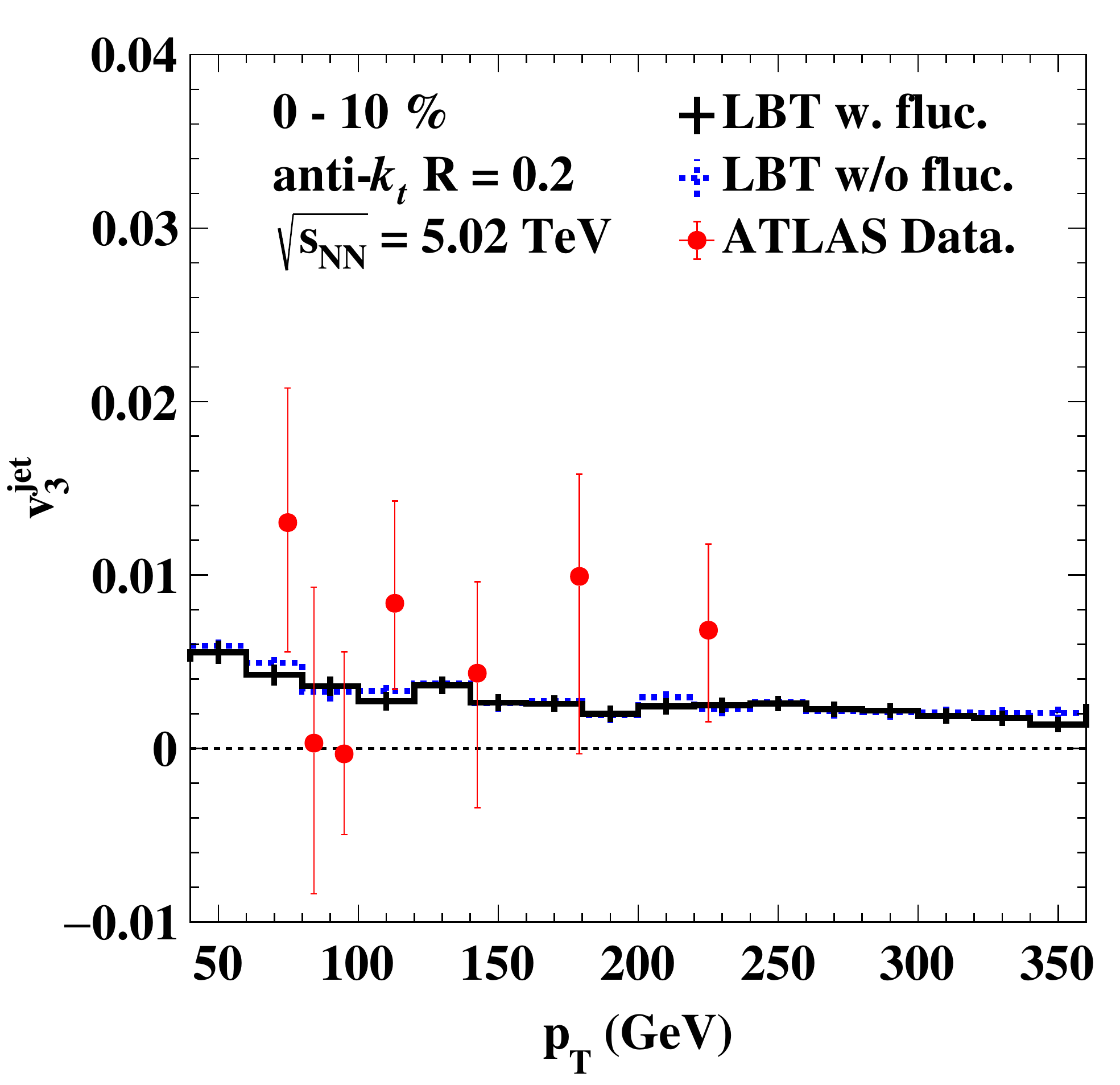}
    \caption{(Color online) The jet $v_{3}^\mathrm{jet}$ as a function of the jet $p_\mathrm{T}$ in 0-10\% Pb+Pb collisions at $\sqrt{s}=5.02$~TeV from the LBT model analyzed with the event plane method (without fluctuation) (dashed line) and scalar product method (with fluctuation) (solid line) as compared to the ATLAS data~\cite{ATLAS:2020qxc}.}
    \label{jetv3_new}
\end{figure}

Shown in Fig.~\ref{jetv3_twoEnergy_withData} is the transverse momentum dependent $v_{3}^\mathrm{jet}$ in Pb+Pb collisions at both $\sqrt{s}=2.76$ and 5.02~TeV, as compared to the ATLAS data. Though the ATLAS data are consistent with LBT results, the experimental errors, mainly statistical, have to be significantly reduced in order to observe the small jet triangular anisotropy. 

To exam the colliding energy and transverse momentum dependence and the effect of bulk flow fluctuations in more detail, we present the same LBT results in Fig.~\ref{jetv3_twoEnergy} with an increased resolution of the vertical axis. We observe the same colliding energy dependence as discussed above. It also decreases slightly with the jet transverse momentum, similar to the $p_{\rm T}$ dependence of $v_2^{\rm jet}$. By comparing the results with bulk flow fluctuations (solid lines) using the scalar product method in Eq.~(\ref{vnjet_Fluc0}) and that without flow fluctuations (dashed lines) using the event plane method Eq.~(\ref{vnjet_noFluc0}), it is, however, difficult to see the effect of the bulk flow fluctuations because of the limited statistics of the LBT simulations in these centrality classes.

To better illustrate the effect of the bulk flow fluctuations, we present $v_{3}^\mathrm{jet}$ from LBT simulations with much higher statistics in Fig.~\ref{jetv3_new} as a function of the jet $p_\mathrm{T}$ in 0-10\% Pb-Pb collisions at $\sqrt{s}=5.02$~TeV. Similar to the conclusion drawn for $v_2^\mathrm{jet}$, the event plane method and the scalar product method lead to very similar values of the jet $v_3^\mathrm{jet}$, indicating negligible effect of the event-by-event bulk flow fluctuations. The results from LBT calculations are consistent with the more recent ATLAS data~\cite{ATLAS:2020qxc} with smaller statistic errors. The small $v_3^\mathrm{jet}$ from LBT is clearly seen to decrease with jet $p_\mathrm{T}$, similar to $v_2^\mathrm{jet}$ shown in Fig.~\ref{jetv2_new}.

\section{Effect of medium response on $v^\mathrm{jet}_2$}
\label{sec:mediumResponse}


\begin{figure}[tbp]
    \includegraphics[width=8cm]{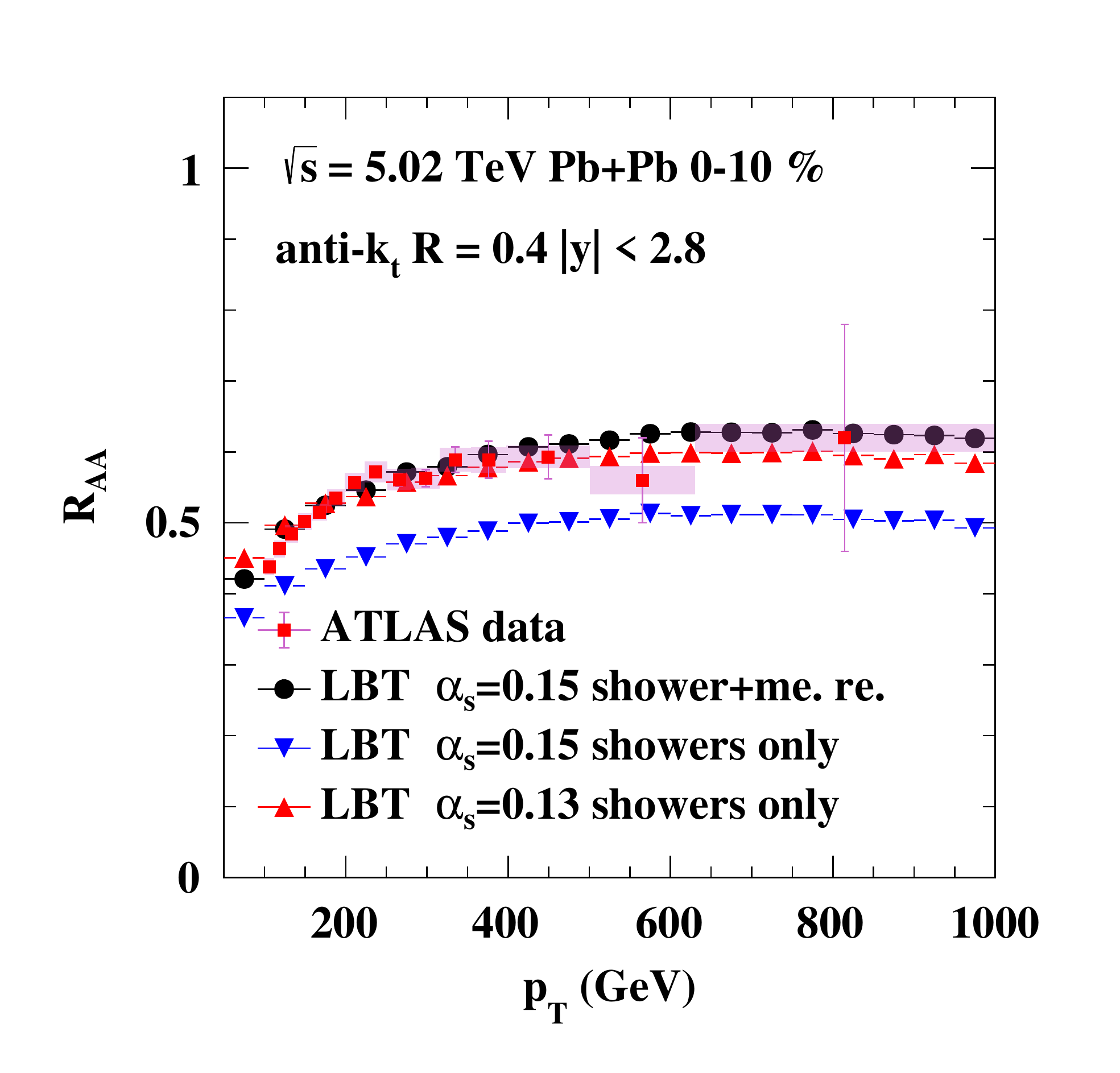}
    \caption{(Color online) Suppression factors of single inclusive jets ($R=0.4$) $R_{\rm AA}$ in 0-10\% Pb+Pb collisions at $\sqrt{s}=5.02$~TeV from LBT with (black circle) and without (blue triangle down) jet-induced medium response for $\alpha_{\rm s} = 0.15$ or without jet-induced medium response but for $\alpha_{\rm s} = 0.13$ (red triangle up), as compared to the ATLAS data~\cite{Aaboud:2018twu}.}
    \label{RAA_R4_alphas}
\end{figure}

\begin{figure}[tbp]
    \includegraphics[width=8cm]{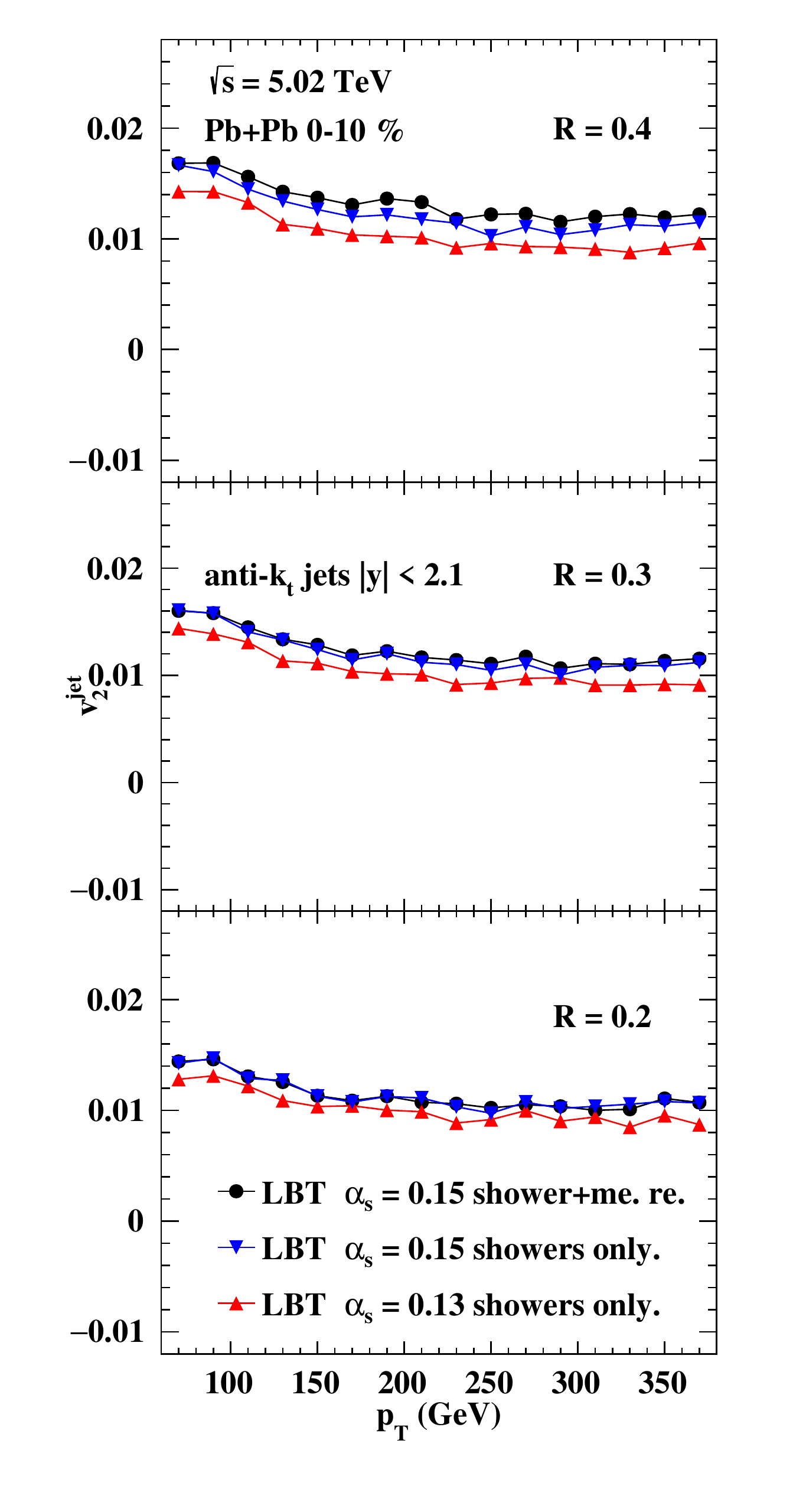}
    \caption{(Color online) Inclusive jet azimuthal anisotropy $v^{\rm jet}_{2}$ in 0-10\% Pb+Pb collisions at $\sqrt{s}=5.02$~TeV with jet cone size $R$ = 0.2, 0.3, 0.4 in mid-rapidity range calculated within the LBT model 
    with (black circle) and without (blue triangle down) jet-induced medium response for $\alpha_{\rm s} = 0.15$ or without jet-induced medium response but for $\alpha_{\rm s} = 0.13$ (red triangle up).  }
    \label{v2pt_R_alphas}
\end{figure}


\begin{figure}[tbp]
    \includegraphics[width=8cm]{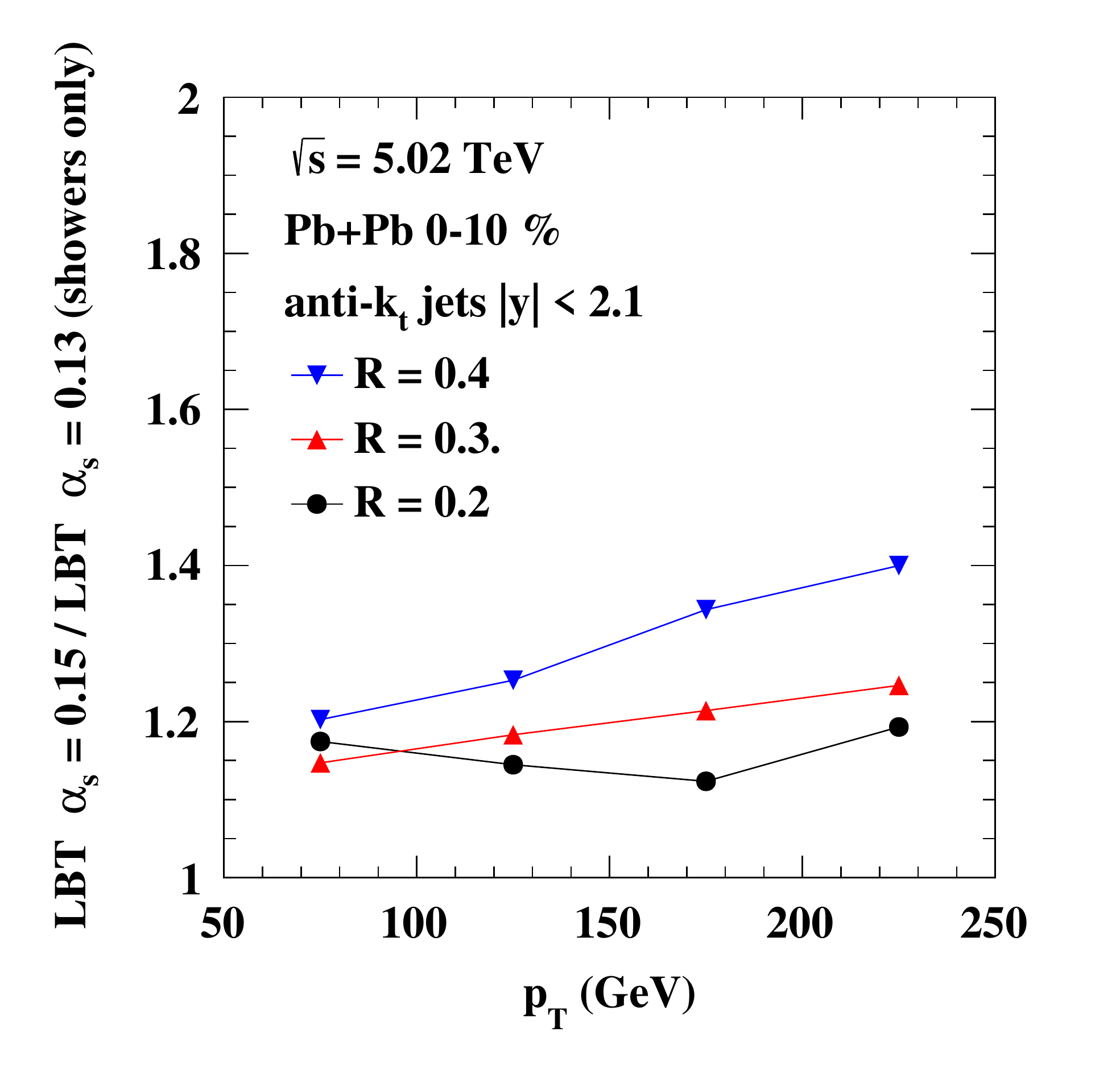}
    \caption{(Color online) The ratio of inclusive $v^{\rm jet}_{2}$ between with medium response effect (at $\alpha_{s} = 0.15$) and without medium response effect (at $\alpha_{s} = 0.13$) in the 0-10\% centrality and mid-rapidity range of 5.02~TeV Pb+Pb collisions  with jet cone size $R = 0.2$ (black circle), 0.3 (blue triangle down) and 0.4 (red triangle up).  }
    \label{v2pt_Rratio}
\end{figure}

One important motivation for developing the LBT model is to investigate the propagation of recoil medium partons and the back-reaction partons on the same footing as the jet shower partons (both leading jet shower partons and radiated gluons induced by jet-medium interaction). Since some of the final hadrons from the hadronization of the recoil medium partons can still fall inside the jet-cone, they will contribute to the total jet energy within the jet cone. The back-reaction induced by the parton transport essentially depletes the phase-space of medium partons behind the propagating jet which is often referred to as the diffusion wake. The energy of these back-reaction medium, or ``negative partons'' within the jet-cone has to be subtracted from the total jet energy. We generally refer recoil and back-reaction or ``negative'' partons as jet-induced medium response.

Effects of jet-induced medium response on net jet energy loss and jet suppression within the LBT and the CoLBT-hydro model have been discussed in detail in previous studies on single inclusive jets~\cite{He:2018xjv}, $\gamma/Z_0$-jets~\cite{Luo:2018pto,Zhang:2018urd}, as well as $\gamma/Z_0$-hadron correlations~\cite{Chen:2017zte,Yang:2021iib} and $\gamma$-jet fragmentation functions \cite{Chen:2020tbl} in heavy-ion collisions. Effects of jet-induced medium response have also been studied within other models such as the MARTINI \cite{Schenke:2009gb,Park:2018acg}, JEWEL \cite{Zapp:2012ak,Zapp:2013vla}, Hybrid \cite{Casalderrey-Solana:2014bpa,Hulcher:2017cpt} and the coupled jet-fluid model \cite{Tachibana:2017syd}. For a recent review on jet quenching and jet-induced medium response see Ref.~\cite{Cao:2020wlm}.  The jet-induced medium response is found to reduce the net jet energy loss and therefore the jet suppression while it enhances soft hadron production both toward the outer edge and outside of the jet cone. It is expected that its effects will also depend on the azimuthal angle of the jet propagation relative to the event plane of the bulk medium. It should therefore also influence the jet azimuthal anisotropy.

Since the inclusion of jet-induced medium response in the jet reconstruction reduces the net jet energy loss, it is also expected to reduce the jet suppression as well as the jet azimuthal anisotropies. Jet energy loss and suppression are generally positively correlated with jet anisotropies. To examine the effect of jet-induced medium response on jet anisotropies beyond such a trivial correlation with the net jet energy loss, we adjust the effective strong coupling constant $\alpha_{\rm s}$ so that the net jet energy loss and the jet suppression remain the same as in the case when jet-induced medium response is included.  Shown in Fig.~\ref{RAA_R4_alphas} are the jet suppression factors $R_{\rm AA}$ for jet cone size $R=0.4$ from LBT simulations of 0-10\% central Pb+Pb collisions at $\sqrt{s}=5.02$ TeV with  (black circle) and without (blue triangle down) medium response for $\alpha_{\rm s}=0.15$ as compared to the ATLAS experimental data~\cite{Aaboud:2018twu}. Without the medium response, LBT results at this effective strong coupling constant have too much suppression as compared to the experimental data. In order to fit the experimental data without the jet-induced medium response, one can reduce the effective strong coupling constant to $\alpha_{\rm s}=0.13$ (red triangle up).  We show in Fig.~\ref{v2pt_R_alphas} the corresponding $v_2^{\rm jet}$ with (black circle) and without (blue triangle down) medium response with the same effective coupling constant and a reduced effective coupling constant (red triangle up) tuned to fit the jet suppression factor. We can see there are noticeable differences in the jet azimuthal anisotropy when jet-induced medium response is excluded from jet reconstruction even when the effective coupling constant is tuned to fit the overall jet suppression. The difference increases with the jet cone-size. 

To exam the azimuthal dependence of the effect of the jet-induced medium response beyond the simple path length dependence of net jet energy loss, we plot the ratios of the jet elliptic anisotropy from LBT with and without jet-induced anisotropy (but with reduced effective coupling constant so that the jet suppression remains the same) in Fig.~\ref{v2pt_Rratio} for different jet cone sizes.  One can see that the medium response increases $v_2^{\rm jet}$ by about 20$\sim$40\% in the $p_\mathrm{T}$ range of 50$\sim$250 GeV/$c$ as compared to the case without medium response even when LBT is tuned in both cases to give the same averaged jet suppression. The increase is bigger for a larger jet cone-size. 

While the azimuthal dependence of the averaged jet path length is the dominant mechanism for the azimuthal anisotropy of jet suppression, it has the opposite effect on the influence of medium response on jet anisotropy. Long (short) path length causes more (less) jet-induced medium response which leads to more (less) reduction of the net jet energy loss and less (more) jet suppression. Therefore such length dependence of the medium response reduces jet anisotropy. However, the combined effect of radial flow and jet-induced medium response can increase jet anisotropy.
According to the study of jet energy loss in Ref.~\cite{He:2018xjv}, radial flow tends to increase the effect of jet-induced medium response and further reduces the net jet energy loss and leads to less jet suppression. This effect of radial flow is bigger for a larger jet cone size. Consequently, the azimuthal modulation of the radial flow, which gives rise to the bulk anisotropic flow, will increase the jet anisotropy. The increase should be bigger for jets with a larger cone size since they contain more contribution from jet-induced medium excitation. These are exactly what we observe in the LBT calculations as shown in Fig.~\ref{v2pt_Rratio}. The LBT results clearly demonstrate that when jet $R_\mathrm{AA}$ is fixed, incorporating the medium response increases $v_2^{\rm jet}$, and stronger enhancement is obtained for jets with larger cone sizes. This also implies that the effect of jet-induced medium response on the reconstructed jet energy is influenced by the radial flow which changes with the azimuthal angle relative to the event plane. The effect understandably increases with the jet cone-size.

\section{Effect of viscosity on $v_2^{\rm jet}$}
\label{sec:viscosity}

So far the hydrodynamic profiles of the bulk medium we have used in the LBT model to calculate the jet anisotropy were given by the CLVisc simulations for an ideal quark-gluon plasma with zero shear viscosity $\eta=0$. However, the shear viscosity of QGP is known to be critical for a more realistic description of the anisotropic flows of the bulk medium with given initial conditions. It is therefore also important to check the effect of the viscosity of the bulk medium on jet quenching, jet anisotropy and hard-soft correlation. 

\begin{figure}[tbp]
    \includegraphics[width=8cm]{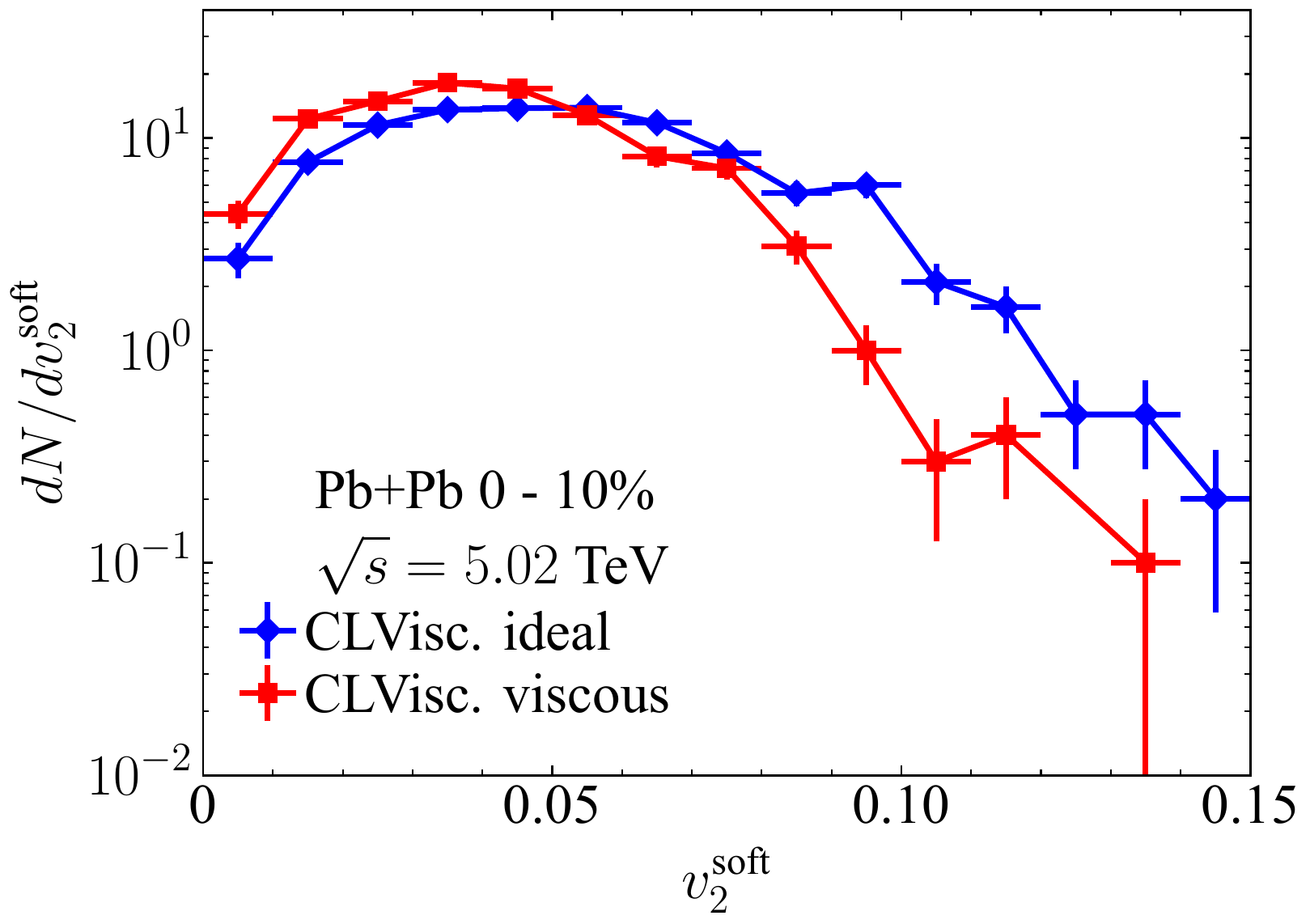}
    \caption{(Color online) Distributions of soft hadron $v^\mathrm{soft}_2$ ($0.3 < p_T<3$ GeV/$c$) from CLVisc model simulations of 0-10\% central Pb+Pb collisions at $\sqrt{s} = 5.02$ TeV with shear viscosity to entropy density ratio $\eta/s=0.08$ (red square) and $\eta/s=0$ (blue diamonds). Errors are statistical with 1000 hydro events.} 
    \label{v2_viscous_5020_0010}
\end{figure}

We carry out the same simulations and jet analyses within LBT in which the space-time profiles of the bulk medium are given by CLVisc hydrodynamic model with the same initial conditions but with finite shear viscosity $\eta/s = 0.08$. The overall scale factor $K$ in the initial conditions for the hydrodynamics in Eq.~(\ref{eq:Pmu}) is adjusted so that the final charged hadron rapidity density in the most central collisions remains the same in ideal and viscous hydrodynamic calculations as compared to the experimental data. Shown in Fig.~\ref{v2_viscous_5020_0010} are the distributions of soft hadron elliptic anisotropy $v^\mathrm{soft}_2$  in 0-10\% central Pb+Pb collisions at $\sqrt{s} = 5.02$ TeV calculated with CLVisc with shear viscosity to entropy density ratio $\eta/s=0$ (blue diamonds ) and $\eta/s=0.08$ (red square). Within the statistic errors for 1000 hydro events, the $v^\mathrm{soft}_2$ distributions from ideal and viscous CLVisc hydrodynamic evolution are similar, except that both the average value and the tail of fluctuation from the viscous hydro are smaller than that from the ideal hydro as expected. Note that we have imposed a cut in $0.3<p_{\rm T}<3$~GeV/$c$ in this calculation of $v^\mathrm{soft}_2$ as in the experimental analysis \cite{Aad:2013xma,Sirunyan:2017fts} which also influences the $v^\mathrm{soft}_2$ distribution slightly. 

\begin{figure}[tbp]
    \includegraphics[width=8cm]{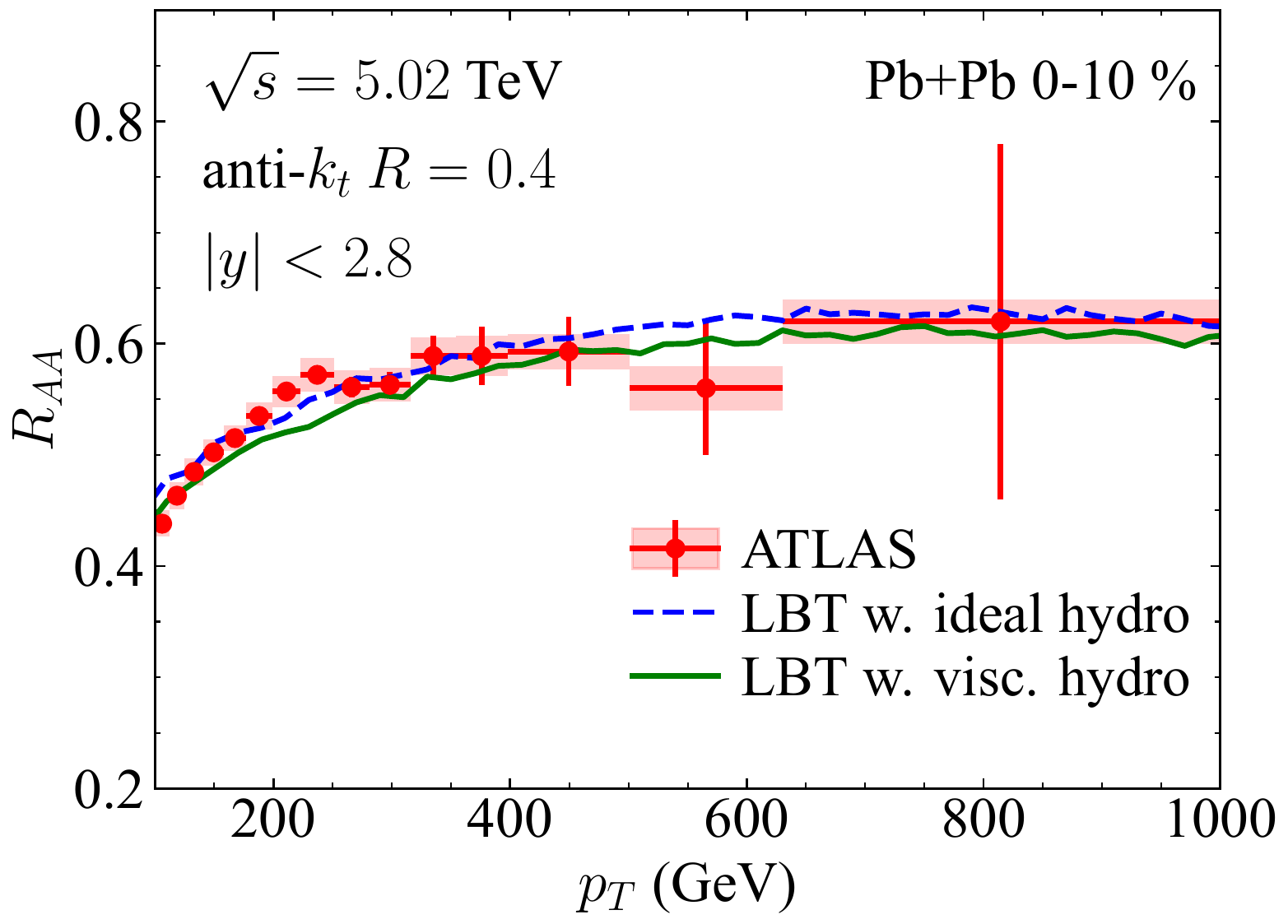}
    \caption{(Color online) Suppression factors of single inclusive jets ($R=0.4$) $R_{\rm AA}$ in 0-10\% Pb+Pb collisions at $\sqrt{s}=5.02$~TeV from LBT with (green solid line) and without (blue dash line) viscosity at $\alpha_{\rm s} = 0.15$, as compared to the ATLAS data~\cite{Aaboud:2018twu}.} 
    \label{RAA_ideal_vs_viscous}
\end{figure}

Using the space-time parton density profiles from these viscous and ideal hydrodynamic events, we first compare in Fig.~\ref{RAA_ideal_vs_viscous} the suppression factors of single inclusive jets (with cone size $R=0.4$) $R_{\rm AA}$ in 0-10\% Pb+Pb collisions at $\sqrt{s}=5.02$~TeV from LBT with viscous (green solid line) and ideal (blue dashed line) hydro profiles with the same effective coupling constant $\alpha_{\rm s} = 0.15$, as compared to the ATLAS data~\cite{Aaboud:2018twu}. Since the entropy density is slightly larger in the viscous hydro than that in the ideal hydro, the corresponding
jet energy  loss is also larger and jets are slightly more suppressed in the viscous hydro than in the ideal hydro. However, the difference is very small.

\begin{figure}[tbp]
    \includegraphics[width=8cm]{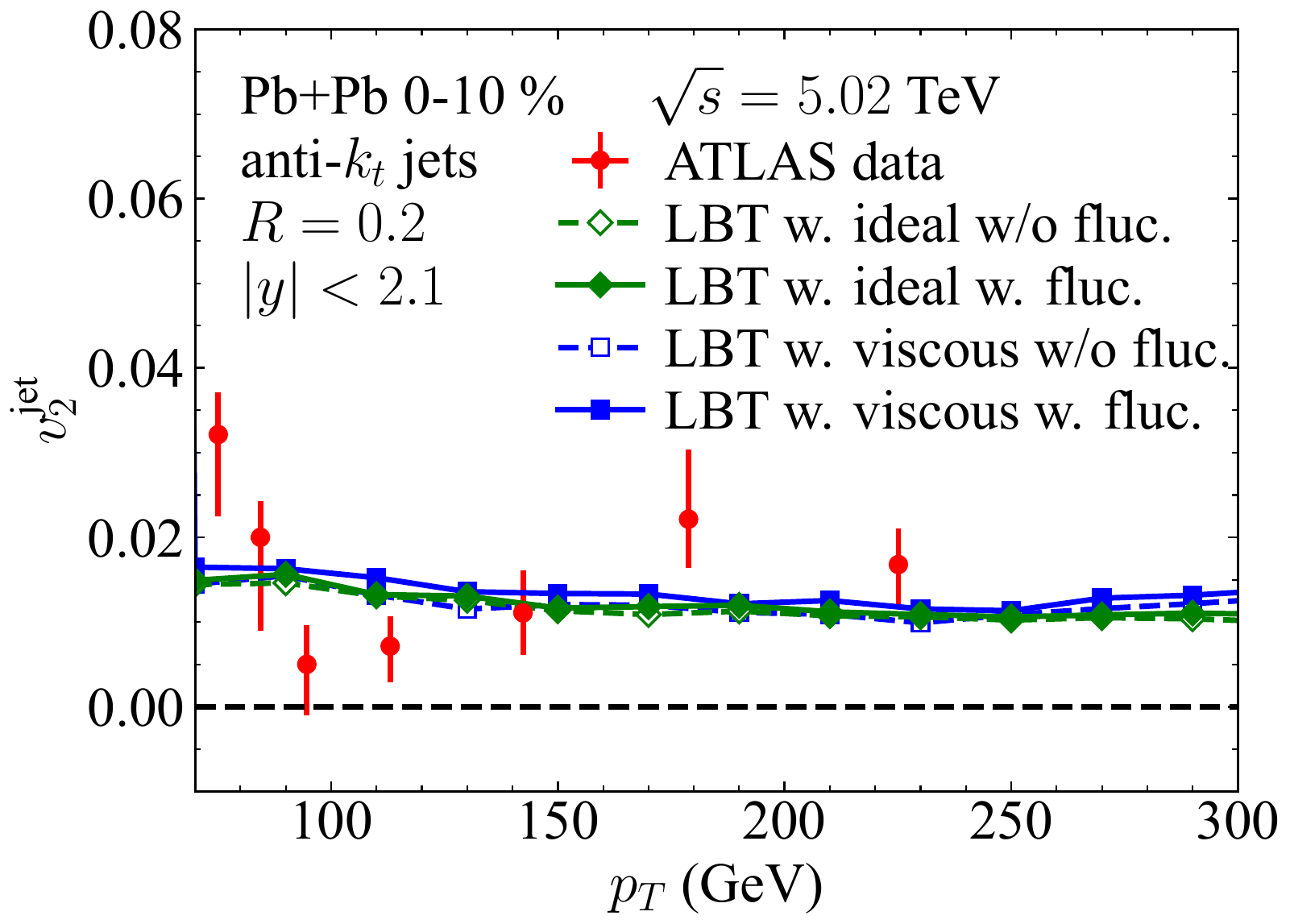}
    \caption{(Color online) The jet elliptic flow coefficient $v_{2}^\mathrm{jet}$ as a function of the jet $p_\mathrm{T}$ from the LBT model calculations with ideal (green diamond) and viscous (blue square) hydro profiles as compared to the ATLAS data~\cite{ATLAS:2020qxc} (filled red circles) in 0-10\% Pb+Pb collisions at $\sqrt{s} = 5.02$~TeV. $v_{2}^\mathrm{jet, SP}$ is labeled as ``with (soft $v_{2}$) fluctuations" (filled symbol and solid line), and $v_{2}^\mathrm{jet, EP}$ is labeled as ``without fluctuations" (open symbol and dashed line).}
    \label{jetv2_ideal_vs_viscous}
\end{figure}

Similarly, we show the jet elliptic flow coefficient $v_{2}^\mathrm{jet}$ as a function of the jet $p_\mathrm{T}$ from LBT model calculations with viscous (blue square-line) and ideal (green diamond-line) hydro profiles as compared to the ATLAS data~\cite{ATLAS:2020qxc} (filled red circles) in 0-10\% Pb+Pb collisions at $\sqrt{s} = 5.02$~TeV in Fig.~\ref{jetv2_ideal_vs_viscous}. $v_{2}^\mathrm{jet, SP}$ is labeled as ``with (soft $v_{2}$) fluctuations" (filled-symbol and solid lines), and $v_{2}^\mathrm{jet, EP}$ is labeled as ``without fluctuations" (open-symbol and dashed lines). One finds that the viscous hydro enhances jet anisotropy slightly, which is mainly due to the larger jet quenching in the viscous hydro than in the ideal hydro, consistent with the effect of viscosity on jet suppression factor $R_\mathrm{AA}$ in Fig.~\ref{RAA_ideal_vs_viscous}.

\begin{figure}[tbp]
    \includegraphics[width=8cm]{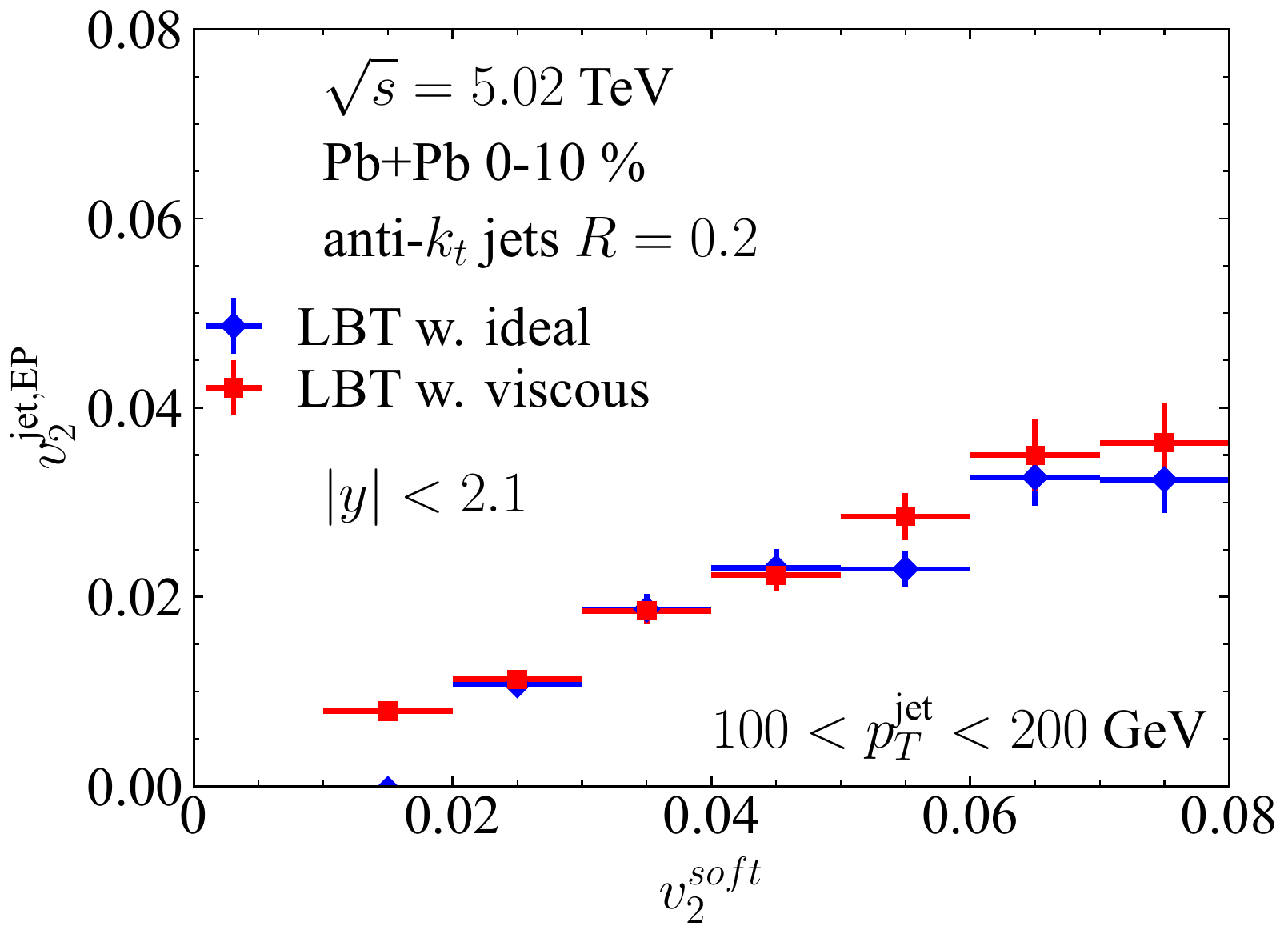}
    \caption{(Color online) Correlations between $v_{2}^\mathrm{jet, EP}$ and bulk $v_2^\mathrm{soft}$ in 0-10\% Pb+Pb collisions $\sqrt{s} = 5.02$~TeV with jet transverse momentum $100 < p_\mathrm{T} < 200$~GeV/$c$ from LBT simulations with viscous (red square) and ideal (blue diamond) hydro profiles.}
    \label{v20jetv2_full_ideal_vs_viscous}
\end{figure}

Finally, we show in Fig.~\ref{v20jetv2_full_ideal_vs_viscous} the correlation between $v_{2}^\mathrm{jet, EP}$ and the bulk $v_2^\mathrm{soft}$ ($0.3 < p_{\rm T}<3$ GeV/$c$) in 0-10\% Pb+Pb collisions at $\sqrt{s} = 5.02$~TeV with jet transverse momentum $100 < p_\mathrm{T} < 200$~GeV/$c$ with viscous (red squares) and ideal (blue diamonds) hydro profiles in LBT. The correlation between jet and soft bulk elliptic flow coefficients in viscous and ideal hydro are almost identical in the region $0.1 < v^\mathrm{soft}_{2} < 0.4$ since the distributions of $v^\mathrm{soft}_2$ in both cases are very similar. In the large region of $v^\mathrm{soft}_2>0.4$, the fluctuation of the bulk $v^\mathrm{soft}_2$ in the viscous hydro is smaller than that in the ideal hydro (see  Fig.~\ref{v2_viscous_5020_0010}). This results in a slightly larger correlation between $v_2^\mathrm{jet, EP}$ and  $v_2^\mathrm{soft}$ in the viscous hydro than that in an ideal hydro.

\section{Summary}
\label{sec:summary}

We have studied in this paper the azimuthal anisotropic coefficients of single inclusive jets produced in relativistic heavy-ion collisions within the LBT model. The AMPT model is used to consistently generate the initial geometric distribution of both the energy density profile for the hydrodynamic evolution of the bulk medium and jet production vertices for jet transport within LBT. The subsequent evolution of the QGP medium is simulated using the CLVisc hydrodynamic model, while the jet-medium interactions are simulated using the LBT model. The only additional model parameter in LBT, the effective strong coupling constant $\alpha_\mathrm{s} = 0.15$, was fixed in our earlier work \cite{He:2018xjv} that provided satisfactory description of the single inclusive jet suppression $R_\mathrm{AA}$ in Pb+Pb collisions at both $\sqrt{s}=2.76$ and 5.02~TeV. Different analysis methods for extracting the anisotropic coefficients of jets have been applied and compared, including the cosine function fit from the azimuthal angular distribution of jets, the event plane method and the scalar product method. Within this framework, we have investigated the transverse momentum dependence, the centrality or participant nucleon number dependence, as well as the colliding energy dependence of the jet anisotropies $v_n^\mathrm{jet}$. Effects of jet-induced medium excitation and viscosity of the bulk medium on the jet $v_2^\mathrm{jet}$ have also been discussed in detail.

We found that, as the centrality increases or the number of participant nucleons decreases, the  $v_2^\mathrm{jet}$ coefficient for jets produced in Pb+Pb collisions at both $\sqrt{s}=2.76$ and 5.02~TeV first increases and then decreases. This non-monotonic behavior, similar to the centrality dependence of $v_2^\mathrm{soft}$ for soft hadrons, results from the competition between the elliptic geometric anisotropy ($\epsilon_2$) of the QGP medium and the amount of net jet energy loss -- the former is larger in more peripheral collisions while the latter is larger in more central collisions. In contrast, the $v_3^\mathrm{jet}$ coefficient appears monotonically decreasing as the centrality increases, for jets produced at both $\sqrt{s}=2.76$ and 5.02~TeV, since the triangular geometric anisotropy ($\epsilon_3$) of the QGP fireball only weakly depends on centrality and $v_3^\mathrm{jet}$ is mainly driven by jet energy loss which decreases monotonically with centrality. Comparing results at the two colliding energies, we noticed that $v_2^\mathrm{jet}$ is larger at $\sqrt{s}=5.02$~TeV than at 2.76~TeV when $N_\mathrm{part}$ is large, but smaller when $N_\mathrm{part}$ is small. This can be understood with the larger $\epsilon_2$ of the QGP profile at $\sqrt{s}=5.02$~TeV than at 2.76~TeV in central collisions, but possible smaller $\epsilon_2$ in peripheral collisions within the AMPT model. In contrast, $v_3^\mathrm{jet}$ remains larger at 5.02~TeV than at 2.76~TeV across the kinematic and centrality region explored in this work. Our study has shown little difference of the $v_2^\mathrm{jet}$ and $v_3^\mathrm{jet}$ coefficients between analyses using the event plane method and the scalar product method, indicating very weak dependence of $v_n^\mathrm{jet}$ on the event-by-event fluctuation of the bulk $v_n^{\rm soft}$. This is in sharp contrast to the conclusion in Ref.~\cite{Noronha-Hostler:2016eow} for high $p_\mathrm{T}$ single inclusive hadrons. To quantify the event-by-event correlation between the jet $v_2^\mathrm{jet}$ and the bulk hadron $v_2^\mathrm{soft}$, we have fitted their correlation function from LBT simulations with a power-law ansatz and found a close to linear dependence of $v_2^\mathrm{jet}$ on  $v_2^\mathrm{soft}$.
Such a linear correlation holds for different colliding energies, centrality classes and jet transverse momenta, and can be tested by more precise jet measurements in the future.

One of the special capabilities of the LBT model is to explore signatures of jet-induced medium excitation in heavy-ion collisions which consists of recoil and ``negative" partons, or Mach wave and diffusion wake, arising from jet-medium interaction. In this work, we found that jet-induced medium response which is influenced by the radial flow increases the jet elliptic anisotropy beyond the simple mechanism of length dependence of jet energy loss. Inclusion of the jet-induced medium response leads to a larger $v_2^\mathrm{jet}$ by 20$\sim$40\% in the $p_\mathrm{T}$ range of 50$\sim$250~GeV/$c$ as compared to the case without medium response but with the same jet suppression factor $R_\mathrm{AA}$ by reducing the effective strong coupling constant. The enhancement of jet $v_2^\mathrm{jet}$ due to jet-induced medium response increases with the jet cone sizes. We also explored the effect of the shear viscosity of the bulk medium on the jet anisotropy which increases $v_2^{\rm jet}$, but only slightly.





\begin{acknowledgments}

We would like to thank Chi Ding for helpful discussions. This work is supported in part by Guangdong Major Project of Basic and Applied Basic Research No. 2020B0301030008, by National Natural Science Foundation of China (NSFC) under Grant Nos. 11935007, 11221504, 11861131009, 11890714, 12175122, 12075098 and 2021-867, by Science and Technology Program of Guangzhou No. 2019050001, by Fundamental Research Funds for Central Universities in China, by the Director, Office of Energy Research, Office of High Energy and Nuclear Physics, Division of Nuclear Physics, of the U.S. Department of Energy under  Contract No. DE-AC02-05CH11231, by the US National Science Foundation under Grant No. ACI-1550228 within the JETSCAPE and OAC-2004571 within the X-SCAPE Collaboration, by EU ERDF and H2020 grant 82409, ERC grant ERC-2018-ADG-835105, Spanish AEI grant FPA2017-83814-P and MDM- 2016-0692, Xunta de Galicia Research Center accreditation 2019-2022. Computations in this study are performed at the NSC3/CCNU and the National Energy Research Scientific Computing Center (NERSC), a U.S. Department of Energy Office of Science User Facility located at Lawrence Berkeley National Laboratory and operated under Contract No. DE-AC02-05CH11231.

\end{acknowledgments}

\bibliography{SCrefs,newrefs}

\end{document}